\documentclass[usenatbib]{mn2e}
\usepackage{longtable}
\input psfig.sty

\title[Binary Population Synthesis]
{Inclusion of Binaries in Evolutionary Population Synthesis}

\author[F. Zhang et al.]
{Fenghui~Zhang,$^{1}$\thanks{E-mail: gssephd@public.km.yn.cn or
zhang\_fh@hotmail.com}
Zhanwen~Han,$^{1}$ Lifang~Li$^{1}$ and Jarrod R. Hurley$^{2}$ \\
$^{1}$ National Astronomical Observatories/Yunnan Observatory,
Chinese Academy of Sciences, PO Box 110, Kunming, \\ Yunnan Province,
650011, China \\
$^{2}$ Centre for Stellar and Planetary Astrophysics, School of
Mathematical Sciences, Monash University, VIC 3800, Australia}

\begin{document}

\date{\today}

\pagerange{\pageref{firstpage}--\pageref{lastpage}}

\pubyear{2004}

\maketitle

\label{firstpage}

\begin{abstract}
Using evolutionary population synthesis we present integrated
colours, integrated spectral energy distributions and
absorption-line indices defined by the Lick Observatory image
dissector scanner (referred to as Lick/IDS) system, for an
extensive set of instantaneous burst binary stellar populations
with and without binary interactions. The ages of the populations
are in the range $1-15\,$Gyr and the metallicities are in the
range $0.0001-0.03$. By comparing the results for populations with
and without binary interactions we show that the inclusion of
binary interactions makes the integrated $\rm U-B$, $\rm B-V$,
$\rm V-R$ and $\rm R-I$ colours and all Lick/IDS spectral
absorption indices (except for $\rm H_\beta$) substantially
smaller. In other words binary evolution makes a population appear
bluer. This effect raises the derived age and metallicity of the
population.

We calculate several sets of additional solar-metallicity binary
stellar populations to explore the influence of input parameters
to the binary evolution algorithm (the common-envelope ejection
efficiency and the stellar wind mass-loss rate) on the resulting
integrated colours. We also look at the dependence on the choice
of distribution functions used to generate the initial binary
population. The results show that variations in the choice of
input model parameters and distributions can significantly affect
the results. However, comparing the discrepancies that exist
between the colours of various models, we find that the
differences are less than those produced between the models with
and those without binary interactions. Therefore it is very
necessary to consider binary interactions in order to draw
accurate conclusions from evolutionary population synthesis work.
\end{abstract}


\section{Introduction}
Amongst the distinct methods available for the study of the
integrated light of stellar populations it is the evolutionary
population synthesis (EPS) method (first introduced by Tinsley
1968) that offers the most direct approach for modelling galaxies.
Remarkable progress in this field has been made during the past
decades but the majority of EPS studies have tended to focus
solely on the evolution of single stars
\citep*{bre94,wor94a,vaz96,kur99}.

Both observation and theory tell us that binary stars play a very
important role in the evolution of stellar population. First,
observations show that upwards of 50\% of the stars populating
galaxies are expected to be in binary or higher-order multiple
systems \citep[for example]{duq91,ric94}. Secondly, binary
evolution, if the component stars are close enough to exchange
mass, can drastically alter the evolution path of a star as
expected from single star evolution. Moreover, binary interactions
can also create some important classes of objects such as blue
stragglers (BSs: \citealt{pol94}), and subdwarf B stars (sdBs,
also referred as extreme horizontal branch [EHB] stars:
\citealt{han2002,han2003}). These objects produced by binary
evolution channels, can significantly affect the integrated
spectral energy distribution (ISED) of a population from
ultraviolet (UV) to radio ranges. Therefore it is necessary to
include binary stars in EPS models. However, in the few EPS
studies to date that have accounted for binary evolution, some
only included specialised classes of binaries (for example,
massive close binary evolution: \citealt*{van98,van2000}), while
others were limited to studying specialised stellar populations
(e.g. \citealt*{pol94,cer97,van97,van2003}). The EPS study of
\citet[hereafter Paper I]{zha2004b} took into account various
known classes of binary stars and investigated general populations
but they only considered solar metallicity populations. This paper
expands that study by including a metallicity dependence and by
investigating the effects of binary interactions and model input
parameters on the results. Furthermore, by using $1 \times 10^6$
instead of $2 \times 10^5$ binary systems the error in the
calculations is reduced and the solar metallicity results of Paper
I are superceded.

In this paper we assume that all stars are born in binaries and
born at the same time, i.e. an instantaneous binary stellar
population (BSP). EPS models of instantaneous burst BSPs require
four key ingredients: (i) a library of evolutionary tracks
(including binary stars and single stars) used to calculate
isochrones in the colour-magnitude diagram (CMD); (ii) a library
of stellar spectra adopted in order to derive the ISED, or
magnitudes and colours, in suitable passbands; (iii) a method to
transform spectral information to absorption-line strengths, for
example, the approach of empirical fitting polynomials
\citep{wor94b}; and (iv) assumptions regarding the various
distributions required for initialization of the binary
population, such as the initial mass function (IMF) of the
primaries and the distribution of orbital separations.

The outline of the paper is as follows: we describe our EPS models
and algorithm in Section 2; our results are presented in Section 3
and in Section 4 we investigate the influences of binary
interactions, model input parameters and distribution functions on
the results; and then finally, in Section 5, we give our
conclusions.

\section{Model Description}
The EPS code used here was developed by Zhang and colleagues
\citep{zha2002,zha2004a,zha2004b} and is based on the most updated
input physics. Here we briefly describe the important features and
components of the model as well as providing details of the
algorithm for computing a BSP.

\subsection{Initialization of the binary population}
To investigate the effect of binary interactions on the EPS study
of stellar populations we first need to construct the
instantaneous burst BSPs. The evolution path taken by a star in a
binary depends critically on the mass of the companion star and
the orbital parameters. As such, binary evolution can be very
complex and it is difficult to estimate the evolutionary
timescales of the two stars without actually evolving the system.
Therefore, the traditional method of constructing populations used
in the EPS study of single stellar populations, where only the IMF
need be considered and the evolutionary timescale of a star is set
as soon as its mass is set, confronts difficulty in the study of
BSPs. Here we need to use a Monte Carlo process, which utilises a
random-number generator in combination with distribution
functions, to generate the initial conditions of a set of binaries
which then need to be evolved using an appropriate binary
evolution model.

To generate a BSP the following input distributions are required
to define the initial state of each binary: (i) the IMF of the
primaries, which gives the relative number of the primaries in the
mass range $M \rightarrow M+{\rm d}M$; (ii) the secondary-mass
distribution; (iii) the distribution of orbital separations (or
periods); and (iv) the eccentricity distribution. For
distributions for which the majority of studies have reached fair
agreement, such as the primary-mass and orbital separation
distributions, we use a certain form while for those that are less
constrained (secondary-mass and eccentricity) we use several
reasonable assumptions. We also need to set the lower and upper
mass cut-offs $M_{{\rm l}}$ and $M_{{\rm u}}$ to the mass
distributions and assign a metallicity $Z$ to the stars.

For each binary system the initial mass of the primary is chosen
from the approximation to the IMF of \cite{mil79} as given by
\citet[][hereinafter EFT]{egg89},
\begin{equation}
M_1 = {\frac{ 0.19X }{(1-X)^{0.75}+0.032(1-X)^{0.25}}} \, ,
\label{mdis}
\end{equation}
where $X$ is a random variable uniformly distributed in the range
[0,1], and $M_1$ is the primary mass in units of ${\rm M_\odot}$.

In choosing the initial mass of the secondary star we can assume
that the masses of the component stars are either correlated or
that they are independent. For the correlated case, the
secondary-mass distribution depends on the primary-mass (as set by
equation \ref{mdis}) and the mass-ratio, $q$, distribution. The
form of the latter is somewhat uncertain and a matter for debate.
In this study we consider two versions of the $q$ distribution,
one is a uniform distribution \citep[EFT 1989;][]{maz92,gol94},
\begin{equation}
n(q) = 1,  \ \ \ \ \     0 \leq q \leq 1,
\label{qdis1}
\end{equation}
where $q = M_2/M_1$, and the other is a thermal distribution
\citep*{han95},
\begin{equation}
n(q) = 2q,  \ \ \ \ \    0 \leq q \leq 1.
\label{qdis2}
\end{equation}
In the uncorrelated case, the secondary mass is chosen
independently from the same IMF as the primary (equation
\ref{mdis}).

The distribution of orbital separations is taken as constant in
$\log a$ (where $a$ is the separation) for wide binaries and falls
off smoothly at close separations:
\begin{equation}
a n(a) = \Bigl\{\matrix{\ a_{\rm sep}(a/a_0)^m , & \ \ \ a \leq
a_0, \cr a_{\rm sep}, \ \ \ \ \ \ \ \ \ & \ \ \ \ \ \ \ \ \ \  a_0
< a < a_1, \cr}
\label{adis}
\end{equation}
where $a_{\rm sep}\approx 0.070$, $a_0=10 {\rm R_{\odot}}$,
$a_1=5.75\times 10^6 {\rm R_{\odot}}$ and $m\approx 1.2$. This
distribution implies that there are equal numbers of wide binary
systems per logarithmic interval, and that approximately 50\% of
the binary systems have orbital periods less than 100\,yr. The
value of 100\,yr can be viewed as the upper limit for interaction
between the component stars -- if the period is longer than
100\,yr the evolution of the stars does not differ from that of
two independent single stars. This fraction of 50\% for binaries
with a period less than 100\,yr is a typical value for the Galaxy,
resulting in $\sim 10\%$ of the binaries experiencing Roche lobe
overflow (RLOF) during the past 13\,Gyr.

In order to investigate the effect of eccentricity we consider two
eccentricity distributions:
(i) all binaries are initially circular, i.e.
\begin{equation}
e = 0 \, , \label{edis1}
\end{equation}
and (ii) all binaries are formed in eccentric orbits where the
initial eccentricity distribution satisfies a uniform form, i.e.,
\begin{equation}
e = X. \label{edis2}
\end{equation}
Under this second assumption large eccentricities in short-period
orbits are to be excluded on the basis that the stars will crash
into each other at periastron.

Equations (\ref{mdis}) - (\ref{edis2})  can be used to set the
initial state of a binary system: the masses of the component
stars, $M_1$ and $M_2$, separation, $a$, and eccentricity, $e$, of
the orbit.

\subsection{Binary evolution model and input parameters}
To describe the evolution of a binary we use the rapid binary star
evolution (BSE) algorithm of \citet*{hur2002}. The BSE algorithm
provides the stellar luminosity $L$, effective temperature $T_{\rm
eff}$, radius $R$, current mass $M$ and the ratio of radius to
Roche-lobe radius $R/R_{\rm L}$ for the component stars, as well
as the period $P$, separation $a$ and eccentricity $e$ for a
binary system. It is valid for component star masses in the range
$0.1 \leq M_1, M_2 \leq 100 M_\odot$, metallicity $0.0001 \leq Z
\leq 0.03$, and eccentricity $0.0 \leq e < 1.0$. The BSE algorithm
includes the single star evolution (SSE) package of analytic
formulae as presented by \citet*{hur2000} in its entirety. In
fact, for orbits that are wide enough that mass exchange between
the component stars does not take place, the evolutionary
parameters of the stars are identical to that given by the SSE
package. The SSE package comprises a set of analytic evolution
functions fitted to the model tracks of \citet{pol98}. Detailed
descriptions of the stellar evolutionary models of \citet{pol98}
and the SSE package have been presented previously in Zhang et al.
(2002: see Sec. 2), so we do not discuss them here. In addition to
all aspects of single star evolution, the BSE algorithm models
processes such as mass transfer, mass accretion, common-envelope
(CE) evolution, collisions, supernova kicks, tidal evolution, and
all angular momentum loss mechanisms. This is done mostly by using
a prescription (or recipe) based approach.

For the BSE code there are several important input parameters that
require mention:
\begin{itemize}
\item (i) the efficiency of CE ejection
$\alpha_{{\rm CE}}$ in the CE evolution model denotes the fraction
of the orbital energy that is transferred to the envelope and is
used to overcome the binding energy. This efficiency is defined
\citep{ibe93} by
\begin{equation}
\alpha_{{\rm CE}} = {\Delta E_{\rm bind} \over \Delta E_{\rm orb}},
\label{cecri}
\end{equation}
where $\Delta E_{\rm orb}$ is the change in the orbital energy of
the binary between the initial and final state of the spiraling-in
process and $\Delta E_{\rm bind}$ is the energy added to the
binding energy of the envelope. If the envelope is ejected before
complete spiral-in then a close binary is the result, otherwise
the two stars will coalesce. This parameter is crucial in
understanding the evolution of populations of binary systems
because it determines the outcome of a CE interaction. In our BSP
study we vary it over a reasonable range ($1.0 - 3.0$) to
investigate its effects.
\item (ii) the coefficient $\eta$ for Reimers' wind mass-loss
\citep{rei75},
\begin{equation}
\dot M_{\rm R} = 4 \times 10^{-13} \eta \frac{LR}{M} M_{\odot}
{\rm yr^{-1}},
\label{wrei}
\end{equation}
where stellar luminosity, $L$, radius, $R$, and mass $M$, are all
in solar units. Equation (\ref{wrei}) gives the wind mass-loss
rate for intermediate and low-mass stars on the giant branch and
beyond. In this study we vary $\eta$ from 0.5 to 0.0.
\item (iii) the tidal enhancement parameter $B$, which appears in the
formula of tidally enhanced mass loss given by \citet{tou88},
\begin{equation}
\dot M = \dot M_{\rm R} \Big[1+B_{\rm W} {\rm max} \big({1 \over
2}, {R \over R_{\rm L}} \big)^6 \Big]
\label{wenh}
\end{equation}
where $\dot M_{\rm R}$ is the Reimers' rate (see equation
\ref{wrei}) and  $R_{\rm L}$ is the Roche lobe radius. Equation
(\ref{wenh}) is used to increase the incidence of RS CVn binaries.
In this study we do not consider tidally enhanced mass loss, i.e.,
$B = 0.0$.
\end{itemize}

\subsection{Stellar spectra and absorption line indices}
The BaSeL-2.0 stellar spectra library of \citet{lej97,lej98}
provides an extensive and homogeneous grid of low-resolution
theoretical flux distributions in the range of $9.1 - 160000\,$nm,
and synthetic UBVRIJHKLM colours for a large range of stellar
parameters: $2000 \leq T_{{\rm eff}}/ {\rm K} \leq 50000$, $-1.02
\leq \log g \leq 5.50$, and $+1.0 \leq {\rm [Fe/H]} \leq -5.0$
(where $g$ denotes surface gravity). For this library correction
functions have been calculated for each value of the $T_{{\rm
eff}}$ and for each wavelength in order to yield synthetic
UBVRIJHKLM colours matching the empirical colour-$T_{{\rm eff}}$
calibrations derived from observations at solar metallicity.
Semi-empirical calibrations for non-solar abundances ($\rm [Fe/H]
= -3.5$ to +1.0) have also been established for this version of
the library. After correction the most important systematic
differences existing between the original model spectra and the
observations are eliminated. Furthermore, synthetic UBV and
Washington ultraviolet excesses $\delta_{\rm (U-B)}$, $\delta_{\rm
(C-M)}$ and $\delta_{\rm (C-T_1)}$, obtained from the original
model spectra of giants and dwarfs, are in excellent agreement
with the empirical metal-abundance calibrations.

The empirical fitting functions of \citet{wor94b} give
absorption-line indices defined by the Lick Observatory image
dissector scanner (Lick/IDS) system as a function of $T_{{\rm
eff}}$, $\log g$, and metallicity [Fe/H]. The effective
temperature spans a range of $2100 \leq T_{{\rm eff}}/{\rm K} \leq
11000$ and the metallicity is in the range $-1.0 \leq {\rm [Fe/H]}
\leq +0.5$. The indices in the Lick system were extracted from the
spectra of 460 stars obtained between 1972 and 1984 using the
red-sensitive IDS and Cassegrain spectrograph on the 3m Shane
telescope at Lick Observatory. The spectra cover the range
$4000-6400$ \AA, with a resolution of $\sim 8$ \AA \
\citep{wor94b}. A more detailed description of the Lick/IDS
spectral absorption-feature indices has been presented in Zhang et
al. (2004a: see Sec. 2).

\subsection{The EPS algorithm}
After choosing a seed for the random number generator we use the
Monte Carlo eqns (\ref{mdis}) - (\ref{edis2}) to produce a
population of $1 \times 10^6$ binary systems. In this work the
lower and upper mass cut-offs $M_{{\rm l}}$ and $M_{{\rm u}}$  are
taken as 0.1 $\rm M_{\odot}$ and 100 $\rm^ M_{\odot}$,
respectively. The relative age, $\tau$, of the BSP is assigned
within the range of $1-15\,$Gyr and the metallicity is chosen
within the limits $0.0001 \le Z \le 0.03$. We can then use the BSE
algorithm to evolve each binary in the BSP to an age of $\tau$
which gives us evolutionary parameters such as $L$, $T_{\rm eff}$,
$R$ and $M$ for the component stars. Next we use the BaSeL-2.0
stellar spectra to transform the evolutionary parameters to
colours and stellar flux and then use the empirical fitting
functions of \citet{wor94b} to derive spectral absorption feature
indices in the Lick/IDS system. Finally, by the following
equations (\ref{inte-c}) - (\ref{inte-mag}) we can obtain the
integrated colours, monochromatic flux and absorption feature
indices for an instantaneous BSP of a particular age and
metallicity. We note that the common-envelope phase of binary
evolution is assumed to be instantaneous in the BSE algorithm and
thus it is not a factor when calculating the flux. Other aspects
of binary evolution that may affect the observed flux, such as the
presence of an accretion disc or novae eruptions, are also not
taken into account, i.e. we only consider the stellar parameters.

In the following equations, a parameter identified by a capital
letter on the left-hand side represents the integrated BSP, while
the corresponding parameter in minuscule on the right-hand side is
for the $k-$th star. The integrated colour is expressed by
\begin{eqnarray}
(C_{{\rm i}}-C_{{\rm j}})_{\tau, Z} & = -2.5 &
\log{\frac{\sum_{{\rm k=1}}^{ {\rm n}}\ 10^{-0.4c_{{\rm i}}}
}{\sum_{{\rm k=1}}^{{\rm n}}\ 10^{-0.4c_{{\rm j}}}}}
\label{inte-c}
\end{eqnarray}
where $c_{{\rm i}}$ and $c_{{\rm j}}$ are the $i$-th and $j$-th
magnitude of the $k$-th star.

The integrated monochromatic flux of a BSP is defined as
\begin{equation}
F_{\lambda,\tau,Z} = \sum_{{\rm k=1}}^{{\rm n}} \ f_{\lambda},
\label{sp-lamda}
\end{equation}
where $f_{\lambda}$ is the SED of the $k$-th star.

The integrated absorption feature index of the Lick/IDS system is
a flux-weighted one. For the $i-$th atomic absorption line, it is
expressed in equivalent width ($W$, in \AA),
\begin{equation}
W_{i,\tau,Z} = {\frac{\sum_{{\rm k=1}}^{{\rm n}} \ w_{{\rm i}} \
\cdot f_{i,{\rm C}\lambda} }{\sum_{{\rm k=1}}^{{\rm n}} \
f_{i,{\rm C}\lambda}}},
\label{inte-EW}
\end{equation}
where $w_{{\rm i}}$ is the equivalent width of the $i-$th index of
the $k$-th star, and $ f_{i,{\rm C} \lambda}$ is the continuum
flux at the midpoint of the $i-$th `feature' passband; and for the
$i-$th molecular line, the feature index is expressed in
magnitude,
\begin{equation}
C_{i,\tau,Z} = -2.5 \log {\frac{\sum_{{\rm k=1}}^{{\rm n}} \
10^{-0.4 c_{{\rm i}}} \cdot \ f_{i,{\rm C}\lambda} }{\sum_{{\rm
k=1}}^{{\rm n}} \ f_{i,{\rm C}\lambda}}}, \label{inte-mag}
\end{equation}
where $c_{{\rm i}}$ is the magnitude of the $i-$th index of the
$k$-th star (as in equation ~\ref{inte-c}).

\section{Results}
\begin{figure}
\psfig{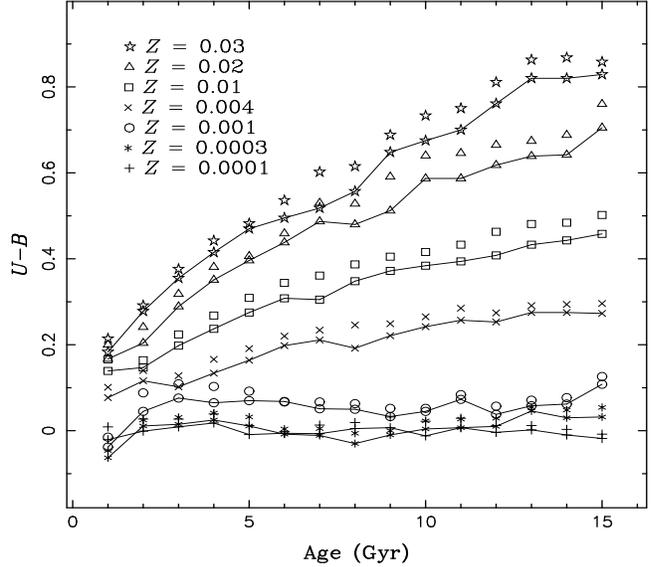}
\caption{The effects of binary interactions on $\rm U-B$ colour
for BSPs of various metallicity. The symbols linked by line are
for BSPs with binary interactions (Model A') and those without a line
are without binary interactions (Model B'). Different symbols
denote different metallicities, from top to bottom, the metallicity
$Z$ is 0.03, 0.02, 0.01, 0.004, 0.001, 0.0003 and 0.0001,
respectively.} \label{ub}
\end{figure}

\begin{figure}
\psfig{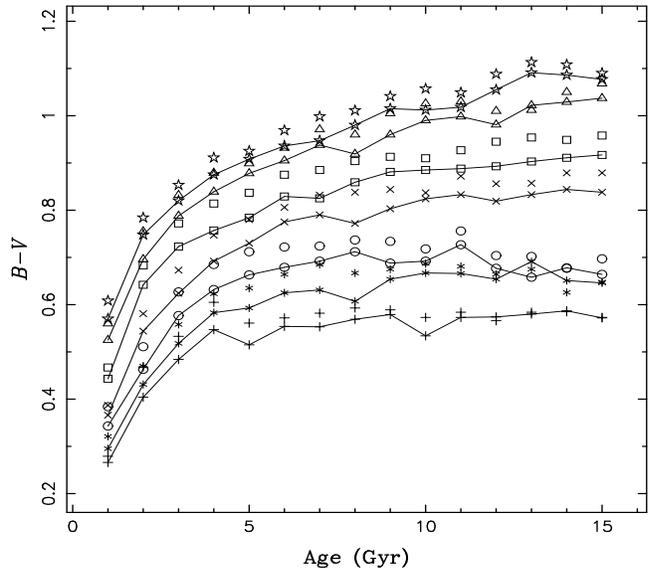}
\caption{Similar to Fig. \ref{ub} but for $\rm B-V$ colour.}
\label{bv}
\end{figure}

\begin{figure}
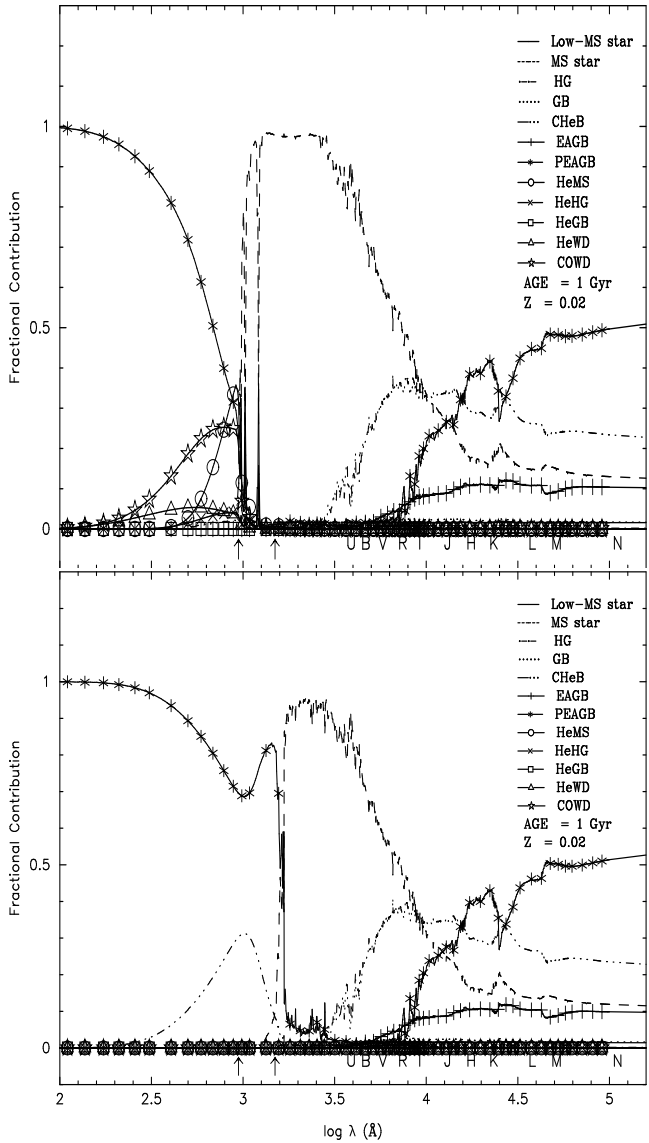

\psfig{file=01-bb2-contri.ps,height=7.5cm,width=8.5cm,bbllx=531pt,bblly=39pt,bburx=80pt,bbury=701pt,clip=,angle=270}
\psfig{file=01-ss2-contri.ps,height=7.58cm,width=8.5cm,bbllx=581pt,bblly=39pt,bburx=80pt,bbury=701pt,clip=,angle=270}
\caption{The fractional contributions of different evolutionary
stages to the total flux are shown for solar metallicity BSPs at
an age of 1 \,Gyr. Top and bottom panels are for Models A' and B',
respectively. The abbreviations are explained in the text.}
\label{contri01}
\end{figure}

\begin{figure}
\psfig{file=13-bb2-contri.ps,height=7.5cm,width=8.5cm,bbllx=531pt,bblly=39pt,bburx=80pt,bbury=701pt,clip=,angle=270}
\psfig{file=13-ss2-contri.ps,height=7.58cm,width=8.5cm,bbllx=581pt,bblly=39pt,bburx=80pt,bbury=701pt,clip=,angle=270}
\caption{Similar to Fig. \ref{contri01} but for an age of 13
\,Gyr.}
\label{contri13}
\end{figure}

\begin{figure}
\psfig{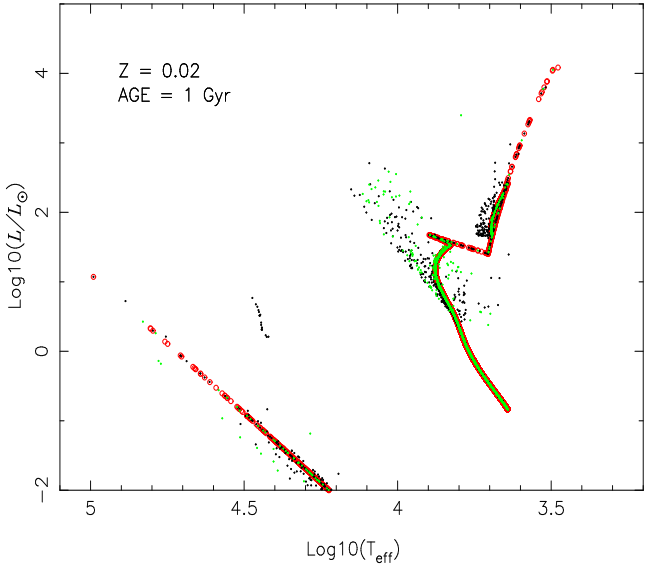}
\caption{Theoretical isochrones for solar-metallicity
instantaneous burst BSPs at an age of 1\,Gyr. Open circles for
Model B', dots for Models A' (black for the primary, blue for the
secondary). For the sake of clarity only $1 \times 10^5$ binary
systems are included and those MS stars with mass $M \la 0.7 {\rm
M_\odot}$ are removed. All binaries are assumed to be resolved.}
\label{syn01}
\end{figure}

\begin{figure}
\psfig{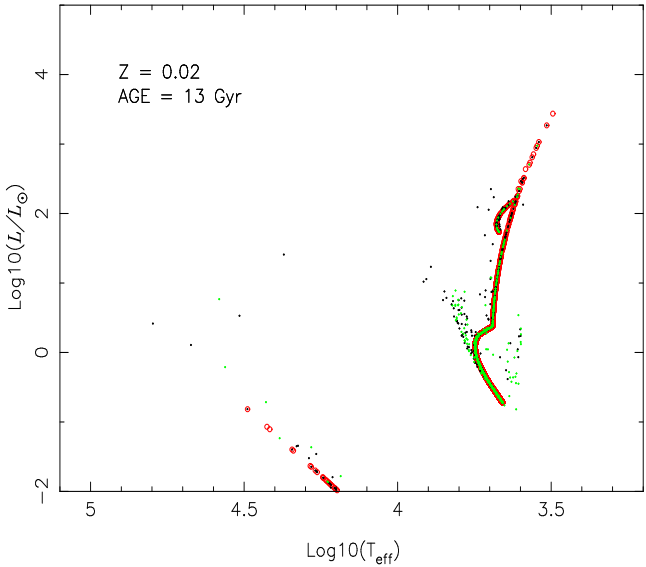}
\caption{Similar to Fig. \ref{syn01} but for an age of 13 \,Gyr.}
\label{syn13}
\end{figure}

\begin{table*}
\centering
\begin{minipage}{1550mm}
\caption{The integrated colours for Model A'.}
\begin{tabular}{rrrrrrrrrrrrrrrr}
\hline \hline
Age  &   1.00  &   2.00  &   3.00  &   4.00  &   5.00  &   6.00  &   7.00  &   8.00  &   9.00  &  10.00  &  11.00  &  12.00  &  13.00  &  14.00  &  15.00  \\
(Gyr) & \multicolumn{15}{c}{} \\
\hline
\multicolumn{16}{c}{$Z$ = 0.0001} \\
$\rm U-B$  &  -0.021  &  -0.001  &   0.009  &   0.018  &  -0.009  &  -0.006  &  -0.007  &   0.005  &   0.007  &  -0.012  &   0.007  &  -0.004  &   0.002  &  -0.010  &  -0.018  \\
$\rm B-V$  &   0.266  &   0.404  &   0.484  &   0.547  &   0.515  &   0.554  &   0.553  &   0.569  &   0.579  &   0.534  &   0.573  &   0.574  &   0.580  &   0.587  &   0.572  \\
$\rm V-R$  &   0.274  &   0.339  &   0.382  &   0.408  &   0.366  &   0.386  &   0.389  &   0.401  &   0.404  &   0.373  &   0.400  &   0.403  &   0.410  &   0.411  &   0.399  \\
$\rm V-I$  &   0.647  &   0.755  &   0.829  &   0.870  &   0.781  &   0.816  &   0.824  &   0.846  &   0.853  &   0.790  &   0.841  &   0.849  &   0.861  &   0.861  &   0.838  \\
\multicolumn{16}{c}{$Z$ = 0.0003} \\
$\rm U-B$  &  -0.063  &   0.011  &   0.015  &   0.025  &   0.011  &  -0.008  &  -0.011  &  -0.030  &  -0.010  &   0.004  &   0.007  &   0.010  &   0.046  &   0.030  &   0.032  \\
$\rm B-V$  &   0.295  &   0.431  &   0.518  &   0.583  &   0.593  &   0.625  &   0.631  &   0.607  &   0.654  &   0.667  &   0.666  &   0.654  &   0.691  &   0.651  &   0.646  \\
$\rm V-R$  &   0.271  &   0.350  &   0.395  &   0.427  &   0.407  &   0.425  &   0.431  &   0.406  &   0.435  &   0.448  &   0.441  &   0.439  &   0.470  &   0.442  &   0.437  \\
$\rm V-I$  &   0.627  &   0.781  &   0.855  &   0.911  &   0.856  &   0.885  &   0.899  &   0.841  &   0.898  &   0.922  &   0.909  &   0.911  &   0.970  &   0.916  &   0.904  \\
\multicolumn{16}{c}{$Z$ = 0.001} \\
$\rm U-B$  &  -0.038  &   0.045  &   0.076  &   0.065  &   0.070  &   0.068  &   0.051  &   0.050  &   0.033  &   0.045  &   0.073  &   0.038  &   0.058  &   0.062  &   0.108  \\
$\rm B-V$  &   0.343  &   0.463  &   0.577  &   0.632  &   0.663  &   0.679  &   0.692  &   0.712  &   0.688  &   0.692  &   0.727  &   0.677  &   0.658  &   0.678  &   0.664  \\
$\rm V-R$  &   0.303  &   0.360  &   0.401  &   0.437  &   0.440  &   0.445  &   0.449  &   0.470  &   0.448  &   0.448  &   0.473  &   0.440  &   0.431  &   0.441  &   0.446  \\
$\rm V-I$  &   0.696  &   0.821  &   0.872  &   0.952  &   0.934  &   0.930  &   0.931  &   0.974  &   0.925  &   0.920  &   0.969  &   0.908  &   0.887  &   0.908  &   0.929  \\
\multicolumn{16}{c}{$Z$ = 0.004} \\
$\rm U-B$  &   0.077  &   0.116  &   0.102  &   0.134  &   0.164  &   0.198  &   0.211  &   0.192  &   0.221  &   0.242  &   0.257  &   0.253  &   0.275  &   0.275  &   0.273  \\
$\rm B-V$  &   0.366  &   0.544  &   0.624  &   0.692  &   0.730  &   0.775  &   0.790  &   0.772  &   0.803  &   0.824  &   0.833  &   0.819  &   0.833  &   0.844  &   0.838  \\
$\rm V-R$  &   0.281  &   0.356  &   0.401  &   0.430  &   0.449  &   0.471  &   0.483  &   0.476  &   0.482  &   0.495  &   0.501  &   0.507  &   0.511  &   0.514  &   0.509  \\
$\rm V-I$  &   0.689  &   0.808  &   0.891  &   0.920  &   0.959  &   0.990  &   1.034  &   1.020  &   1.001  &   1.026  &   1.049  &   1.081  &   1.065  &   1.074  &   1.055  \\
\multicolumn{16}{c}{$Z$ = 0.01} \\
$\rm U-B$  &   0.139  &   0.147  &   0.198  &   0.237  &   0.275  &   0.308  &   0.305  &   0.348  &   0.372  &   0.384  &   0.394  &   0.408  &   0.433  &   0.443  &   0.458  \\
$\rm B-V$  &   0.443  &   0.642  &   0.723  &   0.757  &   0.784  &   0.829  &   0.825  &   0.859  &   0.881  &   0.885  &   0.888  &   0.893  &   0.903  &   0.911  &   0.917  \\
$\rm V-R$  &   0.295  &   0.399  &   0.439  &   0.457  &   0.471  &   0.493  &   0.491  &   0.507  &   0.518  &   0.522  &   0.521  &   0.525  &   0.529  &   0.529  &   0.537  \\
$\rm V-I$  &   0.704  &   0.890  &   0.936  &   0.980  &   0.991  &   1.033  &   1.024  &   1.047  &   1.067  &   1.082  &   1.080  &   1.081  &   1.091  &   1.089  &   1.111  \\
\multicolumn{16}{c}{$Z$ = 0.02} \\
$\rm U-B$  &   0.167  &   0.204  &   0.289  &   0.351  &   0.396  &   0.438  &   0.487  &   0.480  &   0.512  &   0.587  &   0.587  &   0.618  &   0.639  &   0.642  &   0.705  \\
$\rm B-V$  &   0.525  &   0.696  &   0.788  &   0.839  &   0.878  &   0.905  &   0.938  &   0.919  &   0.960  &   0.990  &   0.998  &   0.981  &   1.022  &   1.029  &   1.037  \\
$\rm V-R$  &   0.322  &   0.422  &   0.469  &   0.491  &   0.510  &   0.523  &   0.541  &   0.536  &   0.554  &   0.566  &   0.570  &   0.559  &   0.577  &   0.565  &   0.586  \\
$\rm V-I$  &   0.704  &   0.889  &   0.985  &   1.006  &   1.052  &   1.071  &   1.109  &   1.092  &   1.135  &   1.165  &   1.168  &   1.129  &   1.166  &   1.160  &   1.186  \\
\multicolumn{16}{c}{$Z$ = 0.03} \\
$\rm U-B$  &   0.183  &   0.278  &   0.355  &   0.415  &   0.470  &   0.495  &   0.518  &   0.557  &   0.648  &   0.675  &   0.700  &   0.761  &   0.820  &   0.820  &   0.829  \\
$\rm B-V$  &   0.570  &   0.748  &   0.820  &   0.875  &   0.908  &   0.936  &   0.947  &   0.980  &   1.015  &   1.012  &   1.018  &   1.055  &   1.091  &   1.086  &   1.077  \\
$\rm V-R$  &   0.346  &   0.444  &   0.490  &   0.522  &   0.537  &   0.556  &   0.562  &   0.584  &   0.598  &   0.591  &   0.599  &   0.609  &   0.625  &   0.623  &   0.619  \\
$\rm V-I$  &   0.734  &   0.901  &   1.002  &   1.069  &   1.075  &   1.117  &   1.121  &   1.181  &   1.220  &   1.187  &   1.198  &   1.217  &   1.237  &   1.255  &   1.254  \\
\hline
\end{tabular}
\label{ctab}
\end{minipage}
\end{table*}

In this part we present integrated colours, ISEDs at intermediate
resolution (10 \AA \, in the UV and 20 \AA \, in the visible) and
Lick/IDS absorption feature indices for instantaneous burst BSPs
with and without binary interactions over a large range of age and
metallicity: $1 \leq \tau \leq 15$\,Gyr and $-2.3 \leq {\rm
[Fe/H]} \leq +0.2$. For each model, a total of $1.0 \times 10^6$
binaries are evolved according to the algorithm given in the
previous section.

\subsection{Colours}
In Table \ref{ctab} we present $\rm U-B$, $\rm B-V$, $\rm V-R$ and
$\rm V-I$ colours for the set of BSPs that we call Model A'. In
this model binary interactions are included by utilizing the BSE
algorithm, the initial mass of the secondary $M_2$ is assumed to
be correlated with $M_1$ and is obtained from a uniform mass-ratio
distribution (see equation \ref{qdis1}), eccentric orbits are
allowed for the binary systems and a uniform eccentricity
distribution is adopted (see equation \ref{edis2}), the CE
ejection efficiency $\alpha_{\rm CE}$ is set to 1.0, and the
Reimers' wind mass-loss coefficient $\eta $ is 0.3. The model is
repeated for seven distinct values of Z: 0.03, 0.02, 0.01, 0.004,
0.001, 0.0003 and 0.0001, respectively.

To investigate the effect of binary interactions on the results,
we also construct Model B', which differs from Model A' by
neglecting all binary interactions, i.e. the component stars are
evolved as if in isolation according to the SSE algorithm. In Figs
\ref{ub} and \ref{bv} we give a comparison of the integrated $\rm
U-B$ and $\rm B-V$ colours for Models A' (generated from Table
\ref{ctab}) and B'. We see that the inclusion of binary
interactions makes the integrated $\rm U-B$ and $\rm B-V$ colours
of various metallicity BSPs bluer for all instances.

In Figs \ref{contri01} and \ref{contri13} we give the fractional
contribution of different evolutionary stages to the total flux
for solar metallicity $1-$ and 1$3-$Gyr BSPs, respectively. The
top panels of Figs \ref{contri01} and \ref{contri13} are for
Model A', while the bottom panels are for Model B'. In each of them
various abbreviations are used to denote the evolution phases.
They are as follows: 'MS' stands for main-sequence stars, which
are divided into two phases to distinguish deeply or fully
convective low-mass stars ($M \la 0.7 \rm {M_\odot}$, therefore
Low-MS) and stars of higher mass with little or no convective
envelope ($M \ga 0.7 \rm {M_\odot}$); 'HG' stands for Hertzsprung
gap; 'GB' stands for the first giant branch; 'CHeB' stands for
core helium burning; 'EAGB' stands for early asymptotic giant
branch stars; 'PEAGB' (post EAGB) refers to those phases beyond
the 'EAGB' -- including thermally pulsing giant branch/proto
planetary nebula/planetary nebula (TPAGB/PPN/PN); 'HeMS' stands
for helium star main-sequence; 'HeHG' stands for helium star
Hertzsprung gap; 'HeGB' stands for helium star giant branch; and
'HeWD' and 'COWD' stand for helium and carbon/oxygen white dwarfs,
respectively. Correspondingly, in Figs \ref{syn01} and \ref{syn13}
we plot theoretical isochrones of solar metallicity $1-$ and
$13-$Gyr BSPs. Note that all stars except for
Low-MS stars are included in the isochrones.

Comparing the contribution of different evolutionary stages to the
total flux between Model A' and Model B' for solar metallicity
1-\,Gyr BSPs (Fig. \ref{contri01}), we find that the flux in the
UBVRI passbands is dominated by the same evolutionary stages
regardless of model. This is also true for the $13\,$Gyr solar
metallicity BSPs (Fig. \ref{contri13}). For the young BSPs the
UBVRI flux is dominated by MS stars with mass $M \ga 0.7 {\rm
M_\odot}$, CHeB stars and cooler PEAGB (in fact the TPAGB) stars.
MS and CHeB stars also dominate for the old BSPs but here the GB
stars make a major contribution as well. Therefore, the
differences in the integrated ${\rm U-B}, {\rm B-V}, {\rm V-R}$
and ${\rm V-I}$ colours between Model A' and Model B' will
originate from differences in the distribution of these classes of
stars which dominate the total UBVRI flux. From Figs \ref{syn01}
and \ref{syn13} we see clearly that the distribution of stars in
the Hertzsprung-Russell (HR) diagram for Model A' is significantly
different from that of Model B'. The distribution of the stars is
more dispersed for Model A' in comparison with Model B' and, as
you would expect, BS stars are produced only by Model A'.
Moreover, for young BSPs, Fig. \ref{syn01} shows that Model A'
produces a lot of scatter around the MS  as well as hotter CHeB
stars compared to Model B'. For old BSPs, Fig. \ref{syn13} shows
that Model A' also produces MS, GB and CHeB stars which differ
from the corresponding class of stars in Model B'. The net effect
of the difference in the distribution of these classes of stars is
to make the populations hotter and therefore Model A' appears
younger when compared with Model B'.

So far we have neglected to discuss redder colours (such as, $\rm
J-K$, $\rm H-K$, etc) and the reason is their relatively large
errors. From the top and bottom panels of Fig. \ref{contri01}, we
see that cooler PEAGB stars ($T_{\rm eff} \la 3.54$, see Fig.
\ref{syn01}) give the maximal contribution to the total flux in
those redder passbands for young BSPs. The evolutionary timescale
of these PEAGB stars is so short compared with that of MS stars
that there are few PEAGB stars in our $1 \times 10^6$ binary
system sample. Therefore, for young BSPs large error can result
from fluctuations in the small number of PEAGB stars and thus, the
evolutionary curves of these colours are not smooth. For old BSPs,
from the top and bottom panels of Fig. \ref{contri13} we see that
GB stars contribute about 50\% of the total flux in the redder
wavelength range: in fact, it is GB stars with temperature $T_{\rm
eff} \la 3.54$ that give the maximal contribution. Although the
evolutionary timescale of these stars is longer than that of young
PEAGB stars it is still relatively short compared to the age of
the population and once again a large error can be introduced in
the red flux range. The reason we focus on the bluer colours is
that the total flux in the UVBRI passbands is dominated by stars
which have long evolutionary timescales and the error here is
comparitively small.

\subsection{The integrated spectral energy distribution}
\begin{figure}
\psfig{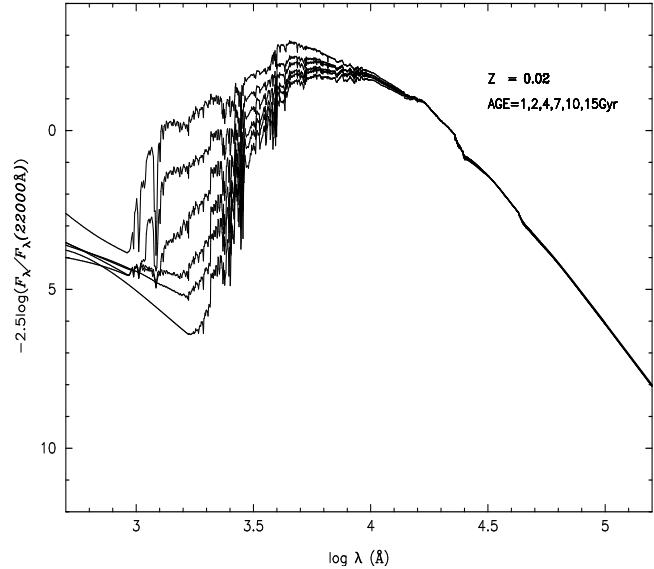}
\caption{The integrated spectral energy distributions as a
function of age for solar-metallicity BSPs with binary
interactions (Model A'). From top to bottom, the ages are
$\tau=1,2,4,7,10$ and 15 \,Gyr, respectively. The flux is
expressed in magnitudes and is normalized to zero at 2.2 $\mu$m.}
\label{ised-t}
\end{figure}

\begin{figure}
\psfig{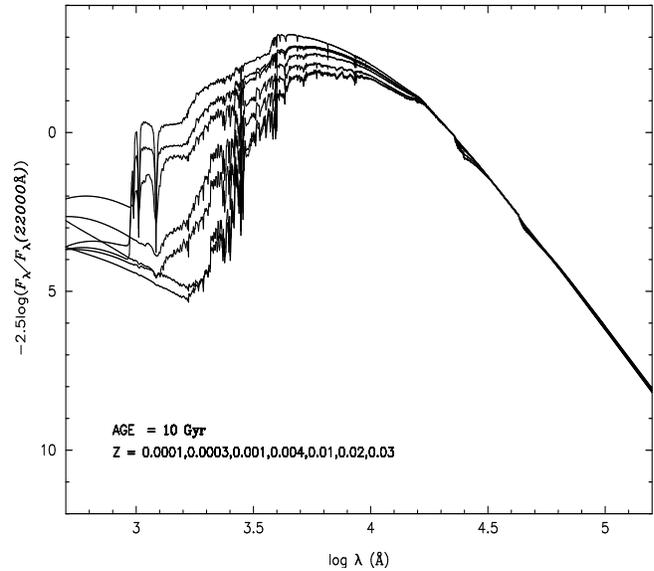}
\caption{The integrated spectral energy distributions as a
function of metallicity for BSPs with binary interactions (Model
A') at age $\tau=10$ \,Gyr. From bottom to top, the metallicity
$Z$ is 0.0001, 0.0003, 0.001, 0.004, 0.01, 0.02 and 0.03,
respectively. The flux is also expressed in magnitudes and is
normalized to zero at 2.2 $\mu$m.}
\label{ised-z}
\end{figure}

\begin{figure}
\psfig{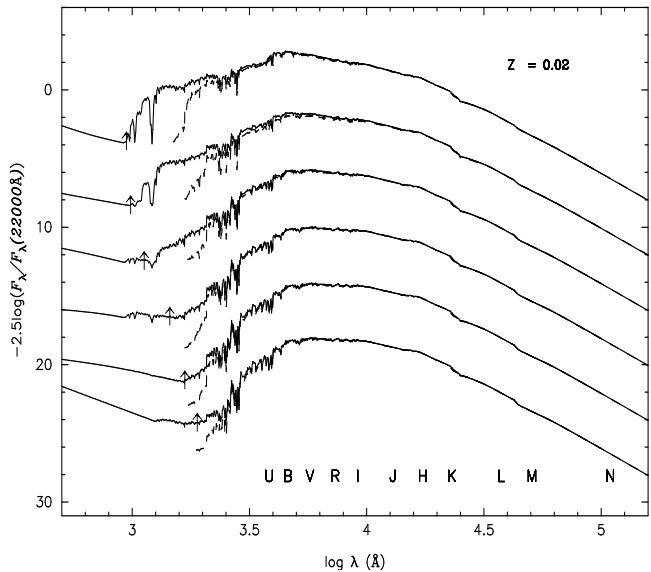}
\caption{The integrated spectral energy distributions for solar
metallicity instantaneous BSPs with (full line) and without
(dashed line) binary interactions at ages $\tau=1,2,4,7,10$ and 15
\,Gyr (from top to bottom, respectively). For Models A' arrows
mark the blue end at which MS stars with mass $M \la 0.7 {\rm
M_\odot}$ contribute 10\% of the ISED, and for Model B' we do not
plot flux shorter than this point. Note for young BSPs the blue
end where MS stars with mass $M \la 0.7 {\rm M_\odot}$ give a 10\%
contribution is significantly different for Models A' and B',
while for old BSPs this discrepancy is insignificant. For the sake
of clarity the fluxes at ages of $2,4,7,10,15$\,Gyr are shifted
downwards by an mount of -4.0.} \label{ised-com}
\end{figure}

In Figs \ref{ised-t} and \ref{ised-z} we give the variation of the
intermediate-resolution ISED with age and metallicity over a wide
wavelength range, $2.7 \leq \log \lambda / {\rm \AA} \leq 5.2$,
for Model A'. The flux is expressed in magnitudes and is
normalized to zero at 2.2 $\mu$m. Figs \ref{ised-t} and
\ref{ised-z} show that the effects of age and metallicity on the
ISEDs are similar, i.e. the ISEDs tend to be redder with
increasing age and metallicity in the wavelength region $3.3 \leq
\log \lambda / {\rm \AA} \leq 4.2$.

In Fig. \ref{ised-com} we compare the ISEDs of Model A' with those
of Model B' for solar-metallicity BSPs at ages $\tau = 1, 2, 4, 7,
10$ and 15\,Gyr (note that the ISEDs for Model A' have also been
given in Fig. \ref{ised-t}). In Fig. \ref{ised-com}, for Model A'
arrows mark the blue end ($\lambda_{1,\tau}$) at which MS stars
with mass $M \ga 0.7 {\rm M_\odot}$ give a 10\% contribution to
the total flux. For Model B' we do not give the fluxes shorter
than the wavelength $\lambda_{2,\tau}$ (defined as for
$\lambda_{1,\tau}$ but for Model B'). Note that for young BSPs
$\lambda_{1,\tau}$ of Model A' is significantly different from
$\lambda_{2,\tau}$ of Model B', with $\lambda_{1,\tau} <
\lambda_{2,\tau}$, while for old BSPs we find $\lambda_{1,\tau}
\approx \lambda_{2,\tau}$. The differences between
$\lambda_{1,\tau}$ and $\lambda_{2,\tau}$ are explained by the
absence of BS stars in Model B', while in Model A' the number of
hot BS stars at early ages is greater than at old ages. Fig.
\ref{ised-com} shows significant disagreement of the ISED between
Model A' and Model B' in the UV and far-UV regions ($\log \lambda
/ {\rm \AA} \la 3.4$): the ISED for Model A' is bluer than that
for Model B' at all ages. Comparing this discrepancy for the young
and old BSPs it seems that it decreases with age: at $\tau=1$\,Gyr
the discrepancy reaches to $\sim$ 5.0 mag at $\log \lambda / {\rm
\AA} = 3.1$ while at $\tau=15$\,Gyr it decrease to $\sim$ 2.2 mag
at $\log \lambda / {\rm \AA} = 3.3$.

To aid investigation of the differences in the ISED in the UV and
far-UV regions between Model A' and Model B' we have included
arrows to mark the wavelength corresponding to $\lambda_{1,\tau}$
(left) and $\lambda_{2,\tau}$ (right) in Figs \ref{contri01} and
\ref{contri13}. For young $\tau=1$\,Gyr BSPs, Fig. \ref{contri01}
shows that the ISED in the region of $\lambda_{2,\tau} \le \log
\lambda / {\rm \AA} \la 3.4$ is dominated by MS stars with mass $M
\ga 0.7 {\rm M_\odot}$ for both Model A' and Model B', and the
fractional contribution of these MS stars drops from 0.93 to 0.1
rapidly at $\lambda_{2,\tau}$ for Model B'. So the discrepancy of
the ISED in this region is mainly introduced by the difference in
the distribution of MS stars with mass $M \ga 0.7{\rm M_\odot}$
(see Fig. \ref{syn01}). In the region $\lambda_{1,\tau} \le
\lambda \le \lambda_{2,\tau}$ the total flux is mainly dominated
by hotter PEAGB and/or CHeB stars for Model B', while still by MS
stars with mass $M \ga 0.7 {\rm M_\odot}$ for Model A', so the
discrepancy in this region is mainly caused by the existence in
Model A' of more hot BSs. This analysis of the $\tau = 1\,$Gyr BSP
is also suitable for the $\tau=2,4,7,10$\,Gyr BSPs. For the old
$\tau=15$\,Gyr BSP, $\lambda_{1,\tau}$ is almost equal to
$\lambda_{2,\tau}$ and in this case the discrepancy of the ISED in
the UV and far-UV regions is caused mainly by the difference in
the distributions of MS stars with mass $M \ga 0.7{\rm M_\odot}$,
hotter PEAGB and/or CHeB stars (as exhibited in Fig. \ref{syn13}).

For the young BSPs ($\tau = 1$ and 2\,Gyr) we also see some
evidence in Fig. \ref{ised-com} of a disagreement of the ISEDs for
Models A' and B' in the visible and infrared regions ($3.5 \la
\log \lambda / {\rm \AA} \la 4.0$). It appears that Model A'
exhibits a bluer continuum than Model B' in this range and we
believe this is owing to fluctuations introduced by the small
number of cool PEAGB stars which, as discussed in Sec. 3.1, can
make a significant contribution to the flux of young populations.

\subsection{Lick spectral absorption feature indices}
\begin{figure*}
\psfig{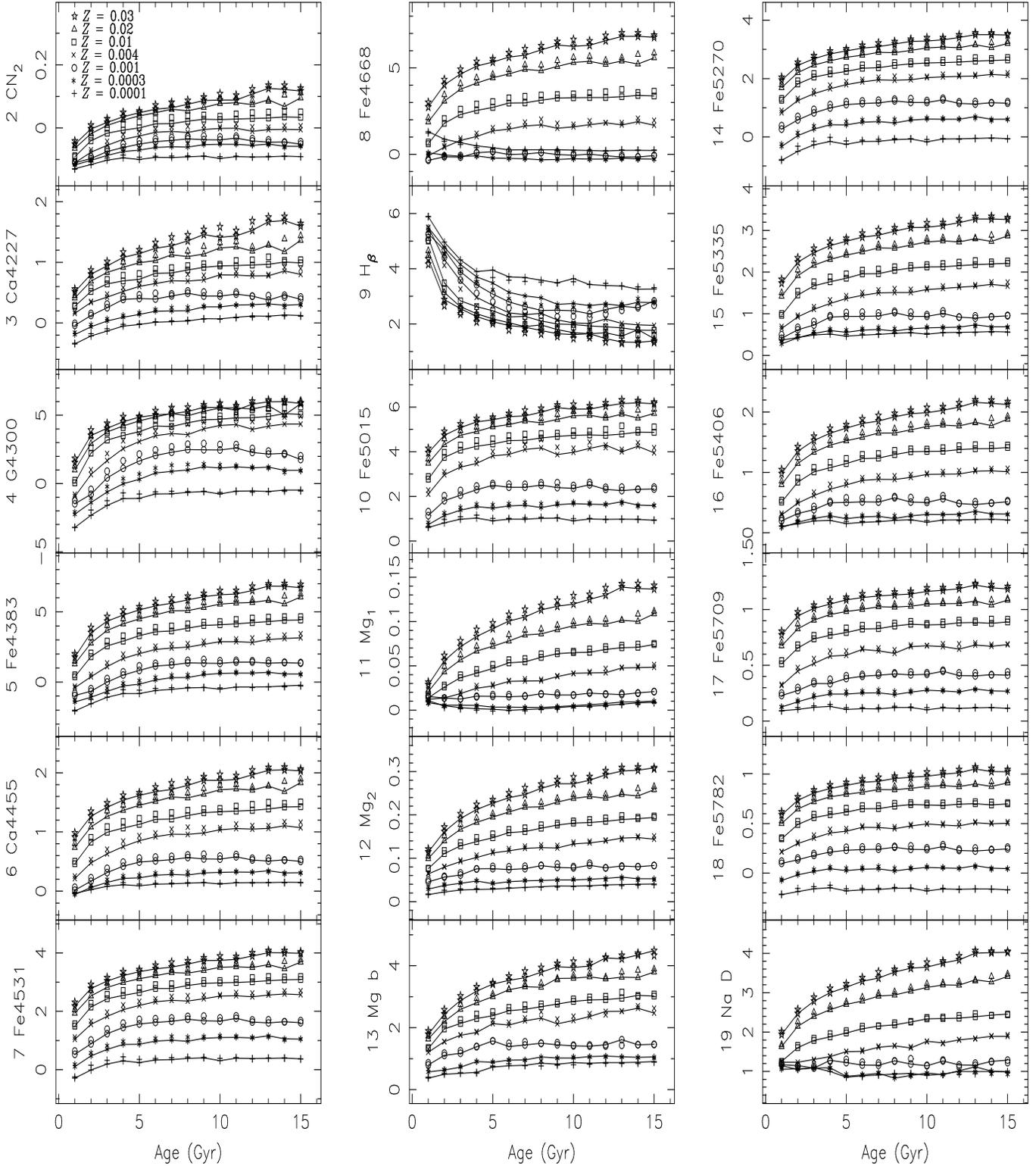}
\caption{The effects of binary interactions on absorption-line
indices in the Lick/IDS system for BSPs of various metallicity.
The symbols linked by a line are for BSPs with binary interactions
(Model A'), those without a line are for BSPs without binary interactions
(Model B'). Different symbol denotes different metallicity, from
top to bottom, the metallicity $Z$ is 0.03, 0..004, 0.001, 0.0003
and 0.0001, respectively.}
\label{lick}
\end{figure*}

In Table \ref{licktab} we present all resulting Lick/IDS spectral
absorption feature indices for Model A' for the seven
metallicities we have considered. In Fig. \ref{lick} we show the
corresponding evolutionary curves. We see that all indices except
for ${\rm H}_\beta$ increase with increasing age and metallicity,
with greater variation at early ages. For comparison, in Fig.
\ref{lick} we also give Lick/IDS indices for Model B'. This
comparison shows that all resulting Lick/IDS indices with binary
interactions are bluer than those without binary interactions and
it appears that these changes in Lick indices due to binaries, are
very small compared to the typical changes with age and
metallicity. The reason that the inclusion of binary interactions
makes all resulting Lick/IDS spectral absorption feature indices
bluer is that all of the indices are within the wavelength region
of 4000 to 6400 \AA, and in this region the ISED for Model A' is
bluer than that for Model B' (see Fig. \ref{ised-com}).

\section{Influence of binary interactions, input parameters/distributions}
\begin{table}
\centering
\begin{minipage}{950mm}
\caption{Model Parameters.}
\begin{tabular}{rrrrrrrrr}
\hline \hline
 Model & BIs & & $e$ & $\alpha_{\rm CE}$ & & $\eta$ & $n(q)$ & $Z$ \\
\hline
A  &  ON       & &   E   &   1.0   & &   0.3   &   CC   & 0.02 \\
B  & {\bf OFF} & &   E   &   1.0   & &   0.3   &   CC   & 0.02 \\
C  &  ON       & &{\bf C}&   1.0   & &   0.3   &   CC   & 0.02 \\
D  &  ON       & &   E   &{\bf 3.0}& &   0.3   &   CC   & 0.02 \\
E1 &  ON       & &   E   &   1.0   & &{\bf 0.0}&   CC   & 0.02 \\
E2 &  ON       & &   E   &   1.0   & &{\bf 0.5}&   CC   & 0.02 \\
F1 &  ON       & &   E   &   1.0   & &   0.3   &{\bf UN} & 0.02 \\
F2 &  ON       & &   E   &   1.0   & &   0.3   &{\bf CQ}& 0.02 \\
\hline
\end{tabular}
\label{mpara}
\end{minipage}
\end{table}

\begin{figure}
\psfig{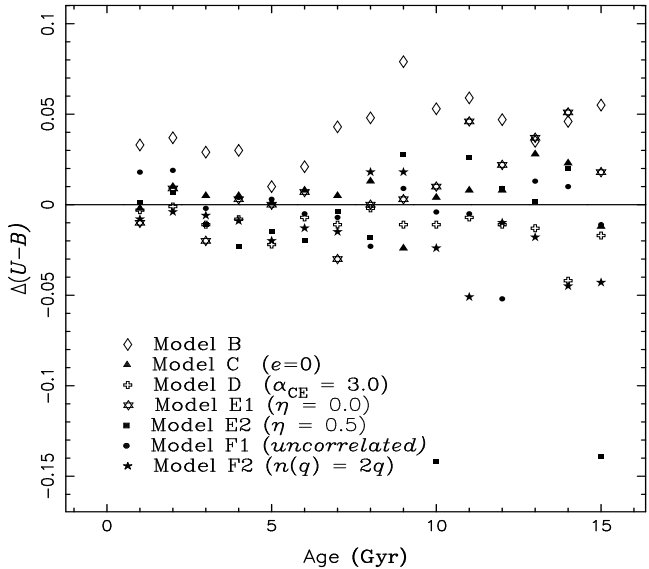}
\caption{Difference of $\rm U-B$ colour between Model A and the other
models (B-F).}
\label{del-ub}
\end{figure}

\begin{figure}
\psfig{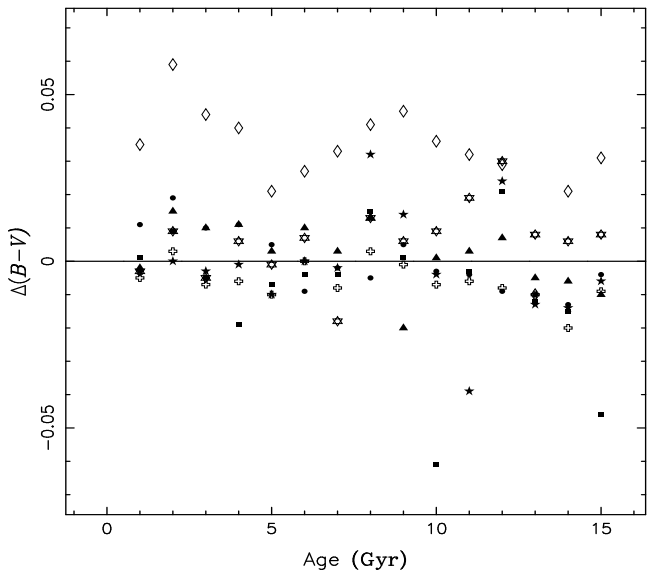}
\caption{Similar to Fig. \ref{del-ub} but for $\rm B-V$ colour.}
\label{del-bv}
\end{figure}

We have performed six additional sets of Monte Carlo simulations
in order to understand the effects of varying the model input
parameters and initial distributions on the resulting U-B and B-V
colours. Here we focus only on solar metallicity BSPs. The
parameters we consider are the CE ejection efficiency $\alpha_{\rm
CE}$ and the Reimers' wind mass-loss coefficient $\eta$. The input
distributions we vary are the eccentricity distribution and the
initial mass distribution of the secondaries.

Each model differs from the other by changing one of the
parameters or distributions used for the initial conditions of the
BSP. The characteristics of each model are summarized in Table
\ref{mpara} where: the {\bf first column} is the name of each
model; the {\bf second column} denotes the condition that binary
interactions (BIs) are taken into account; the {\bf third column}
gives the eccentricity distribution, where 'E' means all binaries
are formed in eccentric orbits and the eccentricity satisfies a
uniform distribution (see equation \ref{edis2}) and 'C' denotes
initially circular orbits (see equation \ref{edis1}); the {\bf
fourth} and {\bf fifth columns} give the values of $\alpha_{\rm
CE}$ and $\eta$; the {\bf sixth} indicates the initial mass
distribution of the secondaries, where 'CC' represents the case
that secondary-mass $M_{\rm 2}$ is correlated with primary-mass
$M_{\rm 1}$ and the mass-ratio distribution $n(q)$ satisfies a
constant form (see equation \ref{qdis1}), 'CQ' represents the
correlated and thermal distribution case (see equation
\ref{qdis2}), and 'UN' is for the uncorrelated case; and the {\bf
last column} gives the metallicity $Z$ (solar in all cases).

Our standard is {\bf Model A} which is Model A' restricted to
solar metallicity. Similarly {\bf Model B} is Model B' restricted
to only solar metallicity. For {\bf Model C} binary interactions
are taken into account but all binary orbits are initially
circular. {\bf Model D} is the same as Model A except that
$\alpha_{\rm CE} = 3.0$. {\bf Model E1} does not include any
stellar wind ($\eta=0.0$) while in {\bf Model E2} $\eta=0.5$. To
investigate the changes produced by altering the initial
secondary-mass distribution {\bf Model F1} chooses the
secondary-mass independently from the same IMF as used for the
primary while in {\bf Model F2} the secondary-mass is correlated
with primary-mass and the mass-ratio satisfies a thermal
distribution.

To discuss the effects of binary interactions and the input
parameters/distributions on the appearance of BSPs we use the
difference of the integrated colours ($\Delta ({\rm U-B})_{j-{\rm
A}}$ and $\Delta ({\rm B-V})_{j-{\rm A}}$) obtained by subtracting
the colour for Model A from that of Model $j$ ($j =\,$B-F). All
results for Model A have been given in Section 3 and the
differences are plotted in Figs \ref{del-ub} and \ref{del-bv}.

\subsection{Binary interaction (Model B)}
Binary stars play a very important role in EPS studies and as we
pointed out earlier the majority of current EPS studies have only
included single star evolution. In Section 3 we have already
discussed the effect of the modelling of binary interactions on
the integrated colours, ISED and Lick/IDS absorption line indices
of a population for a range of metallicities. From the comparison
of the integrated $\rm U-B$ and $\rm B-V$ colours for Models A and
B in Figs \ref{del-ub} and \ref{del-bv} we see that the colours
for Model A are less than those for Model B in almost all
instances (the exception is $\rm B-V$ at 13\,Gyr), i.e. for both
colours considered the solar metallicity BSPs including binary
interactions are bluer than those without binary interactions. For
the full metallicity range ($Z=0.0001, 0.003, 0.001, 0.004, 0.01$
and 0.03) Figs \ref{ub} and \ref{bv} show that binary interactions
play the same effect on $\rm U-B$ and $\rm B-V$ colours. Also Fig.
\ref{lick} shows that binary interactions make all Lick/IDS
absorption line indices bluer for BSPs regardless of metallicity.
The reason that binary interactions make the integrated $\rm U-B$
and $\rm B-V$ colours and the Lick/IDS absorption indices bluer
has been discussed in Section 3 and we do not take it further
here.

\subsection{The distribution of orbital eccentricity  (Model C)}
\begin{figure}
\psfig{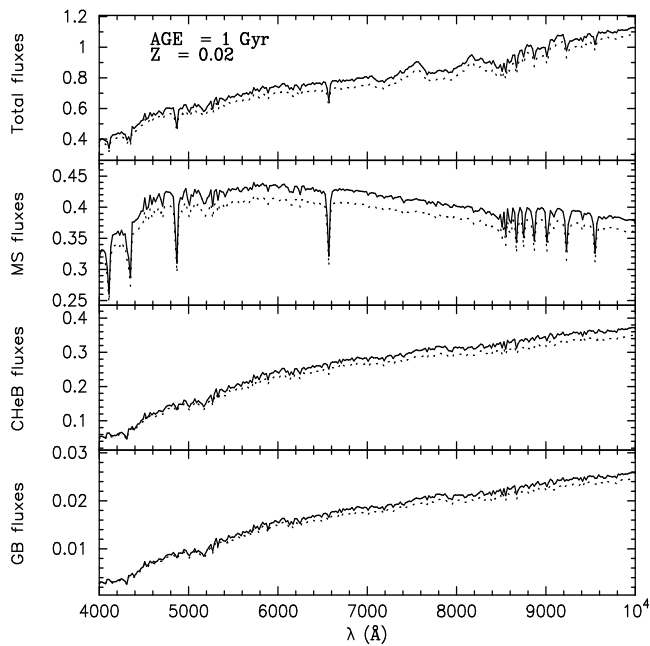}
\caption{The total flux and the the fluxes of all MS stars with
mass $M \ga 0.7 {\rm M_\odot}$, CHeB and GB stars for Models A
(full line) and C (dotted line) at an age of 1\,Gyr. Each flux is
expressed in units of the total flux at 2.2 $\mu$m ($F_{{\rm
\lambda,2.2} \mu {\rm m}}$) for the corresponding model.}
\label{ised-s-ecc01}
\end{figure}

\begin{figure}
\psfig{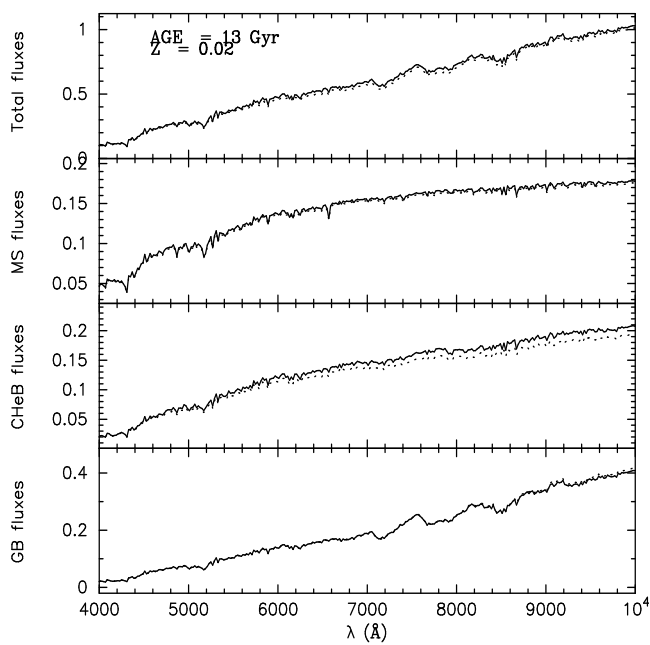}
\caption{Similar to Fig. \ref{ised-s-ecc01} but for an age of 13 \,Gyr.}
\label{ised-s-ecc13}
\end{figure}

Observations of binary stars show that the two components of the
binary often move in eccentric orbits and that there are some
systems for which the eccentricity approaches unity. Thus it would
be wrong to neglect eccentricity when modelling binary evolution
and accordingly \citet{hur2002} included eccentric orbits in their
BSE algorithm. Statistical studies for the distribution of orbital
eccentricity of binaries have found that it appears to be roughly
uniform in the range from zero to unity (see equation
\ref{edis2}), at least for those binaries whose components are
sufficiently well separated that they can have had little chance
yet to interact with each other, even at periastron \citep{egg04}.
So in our standard model (Model A) we assumed that all binaries
are formed with an eccentricity drawn from an uniform eccentricity
distribution. However, to give an idea of how sensitive the
integrated $\rm U-B$ and $\rm B-V$ colours are to changes in this
distribution in Model C we consider the rather drastic case that
all binaries are initially circular initially.

In Figs \ref{del-ub} and \ref{del-bv} we present the differences in
the integrated $\rm U-B$ and $\rm B-V$ colours between Model C and
Model A, i.e. $\Delta (\rm U-B)_{(C-A)}$ and $\Delta (\rm
B-V)_{(C-A)}$. Fig. \ref{del-ub} shows that the integrated $\rm
U-B$ colour for Model C is greater than that for Model A at most
ages (except at $\tau=1,9$ and 15\,Gyr), which means that if all binaries
are in circular orbits the resulting $\rm U-B$ colour is redder,
and thus, the populations appear older. In Fig. \ref{del-bv} a similar
effect on the integrated $\rm B-V$ colour appears except at
$\tau=1,9,13-15$\,Gyr.

Why does Model C make the integrated $\rm U-B$ and $\rm B-V$
colours redder? In the BSE algorithm tidal interactions are
modelled (circularisation and synchronisation) which means that an
eccentric orbit may be circularised, if the stars are close enough
for tides to become strong, and the orbital separation decreased.
Now consider a binary in Model C which has a circular orbit and a
separation $a$ such that mass-transfer is just avoided. The same
binary in Model A with an initial eccentricity will experience
tidal circularisation (most likely when the primary star reaches
the GB) which will decrease $a$ and lead to mass-transfer.
Also, closer systems that would initiate mass-transfer in Model C
but not merge now are likely to lead to coalescence in Model A.
In summary, the variation of the eccentricity distribution not
only can influence the evolutionary path of close binaries but
also that of relatively wide ones.

For binary populations, the variation in evolutionary path of
binaries will cause variations in the number of stars and the
distribution of stars in the CMD for different evolutionary
phases. These two effects determine the discrepancies in the
integrated colours and ISEDs. Because it is extremely difficult to
describe the variation of the distribution of stars in the CMD we
will focus only on the variation of the numbers of MS stars with
mass $M \ga 0.7 {\rm M_\odot}$, CHeB stars and GB stars. The
reason is that the ISED in the UBV passbands is mainly contributed
by these three evolutionary phases for both young (although GB
stars do not contribute so strongly) and old BSPs (see Figs
\ref{contri01} and \ref{contri13}). For $\tau=1$\,Gyr BSPs we find
that: (i) on the MS the number of primaries with mass $M \ga
0.7{\rm M_\odot}$ for Model C is less than that for Model A (i.e.,
$\Delta N_{\rm p,MS,(C-A)}<0$) while that of secondaries with mass
$M \ga 0.7{\rm M_\odot}$ for Model C is greater (i.e., $\Delta
N_{\rm s,MS,(C-A)}>0$) and the decrease in the number of primaries
is less than the increase in the number of secondaries (i.e.,
$\Delta N_{\rm p,MS,(C-A)} + \Delta N_{\rm s,MS,(C-A)}>0$); (ii)
on the GB both the numbers of primaries and secondaries almost do
not vary; and (iii) during CHeB the number of primaries for Model
C is less than that for Model A ($\Delta N_{\rm p,CHeB,(C-A)}<0$)
while the number of secondaries is greater ($\Delta N_{\rm
s,CHeB,(C-A)}>0$) and $\Delta N_{\rm p,CHeB,(C-A)} \approx \Delta
N_{\rm s,CHeB,(C-A)}$. For $\tau=13$\,Gyr BSPs on the MS and GB
both the numbers of primaries and secondaries for Model C are
greater than those for Model A, while the opposite holds during
CHeB.

In Figs \ref{ised-s-ecc01} and \ref{ised-s-ecc13} we give the
total flux and the fluxes of all MS stars with mass $M \ga 0.7
{\rm M_\odot}$, CHeB stars and GB stars over a wavelength range,
$4000 \le \lambda \le 10000 {\rm \AA}$, for Models A and C at ages
of 1\,Gyr and 13\,Gyr, respectively. Each flux curve is expressed
in units of the total flux at 2.2 $\mu$m (i.e., $F_{{\rm
\lambda,2.2} \mu {\rm m}}$) for the corresponding model. Fig.
\ref{ised-s-ecc01} shows that the total flux and the fluxes
eminating from the three evolutionary phases are redder for Model
C than for Model A at an age of $\tau=1$\,Gyr. Comparing the
discrepancies in the fluxes between Models C and Model A we can
conclude that the redder total flux (the top panel of Fig.
\ref{ised-s-ecc01}) and the redder colours (Figs \ref{del-ub} and
\ref{del-bv}) for Model C are mainly introduced by the difference
in MS stars with mass $M \ga 0.7 {\rm M_\odot}$ at an age of
$\tau=1$\,Gyr. For the $\tau=13$\,Gyr BSP, Fig. \ref{ised-s-ecc13}
shows that the total flux and the flux of all CHeB stars for Model
C are redder than for Model A, and the discrepancies in the fluxes
of all MS stars with mass $M \ga 0.7{\rm M_\odot}$ and GB stars is
insignificant. Thus it is the CHeB stars that are influencing the
redder total flux (top panel of Fig. \ref{ised-s-ecc13}) and the
redder colours (Figs \ref{del-ub} and \ref{del-bv}) for Model C at
$13\,$Gyr.

\subsection{The common-envelope ejection efficiency $\alpha_{\rm
CE}$  (Model D)}
\begin{figure}
\psfig{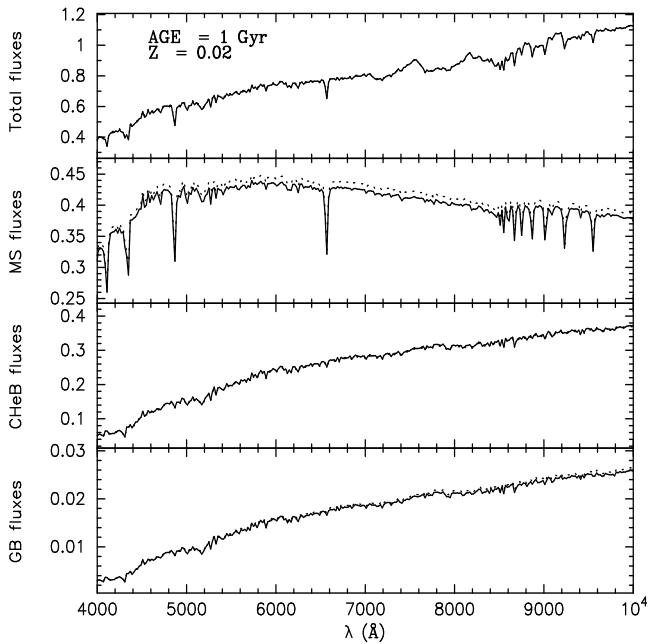}
\caption{The total flux and the the fluxes of all MS stars with
mass $M \ga 0.7 {\rm M_\odot}$, CHeB and GB stars for Models A
(full line) and D (dotted line) at an age of 1\,Gyr. Each flux is
expressed in units of the total flux at 2.2 $\mu$m ($F_{{\rm
\lambda,2.2} \mu {\rm m}}$) for the corresponding model.}
\label{ised-s-ace01}
\end{figure}

\begin{figure}
\psfig{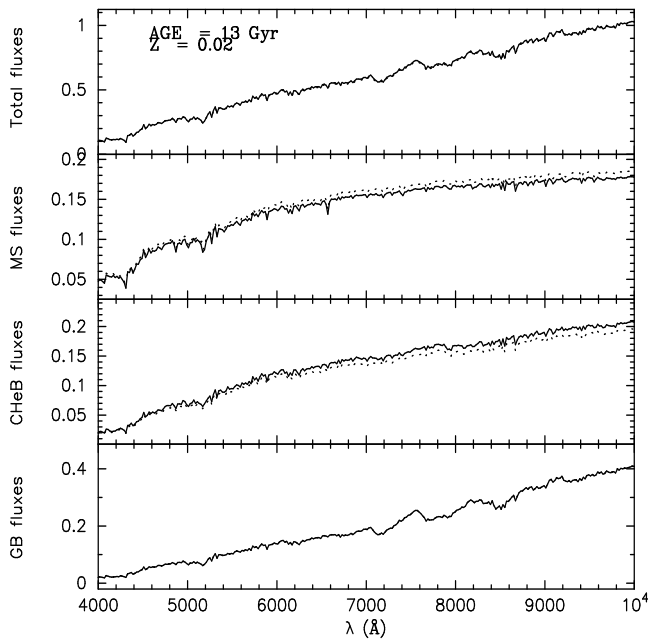}
\caption{Similar to Fig. \ref{ised-s-ace01} but for an age of 13 \,Gyr.}
\label{ised-s-ace13}
\end{figure}

Common-envelope evolution is one of the most important and complex
but also one of the least understood phases of binary evolution.
One area of the uncertainties is related to the criterion for the
ejection of the CE, which crucially determines the orbital period
distribution of post-CE binaries \citep{han2002}. In the BSE code
\citet{hur2002} adopt a widely used and relatively simple
criterion where the CE is ejected when the change in orbital
energy, multiplied by the CE ejection efficiency parameter,
$\alpha_{\rm CE}$, exceeds the binding energy of the envelope (see
equation \ref{cecri}). The approach used by \citet{han2002}
differs in that the thermal energy of the envelope is also
considered in their criterion. The value of $\alpha_{\rm CE}$ is
an uncertain but very crucial factor as it determines the
evolution path of post-CE binaries and therefore affects the birth
rates and numbers of the various types of binary in binary
population synthesis (BPS). For example, using the BSE algorithm
\citet{hur2002} showed that a particular post-CE binary that went
on to form a cataclysmic variable when $\alpha_{\rm CE} = 1$ was
used would emerge from CE evolution with too wide an orbit for
subsequent interaction if $\alpha_{\rm CE} = 3$ was used instead.
Knowledge of the typical values of $\alpha_{\rm CE}$ is therefore
crucial in understanding the evolution of populations of binary
systems. Comparing observational birth rates and numbers of the
various binary types with those obtained by the method of BPS, we
can set constrains on $\alpha_{\rm CE}$ theoretically, but in fact
this method is difficult because selection effects involved in the
observations weaken the constraints. In this work we vary
$\alpha_{\rm CE}$ in a reasonable range to investigate its effect:
in Model A $\alpha_{\rm CE}=1$, while in Model D it is set to 3.

Figs \ref{del-ub} and \ref{del-bv} give the difference in the
integrated $\rm U-B$ and $\rm B-V$ colours between Models D A,
i.e., $\Delta (\rm U-B)_{(D-A)}$ and $\Delta (\rm B-V)_{(D-A)}$.
Fig. \ref{del-ub} shows that the integrated $\rm U-B$ colour for
Model D is less than that for Model A in all instances and Fig.
\ref{del-bv} shows that the integrated $\rm B-V$ colour for Model
C is less than that for Model A except at $\tau = 2$ and $8\,$Gyr.
That is to say, by increasing $\alpha_{\rm CE}$ the resulting $\rm
U-B$ and $\rm B-V$ colours have become bluer and the population
appears younger. The effect of Model D on the integrated $\rm U-B$
and $\rm B-V$ colours is contrary to that of Model C.

Because $\alpha_{\rm CE}$ is introduced in CE evolution its
variation only affects the evolution of binaries experiencing a CE
phase: it affects the CE evolutionary process and the evolution
path of post-CE descendants. If $\alpha_{\rm CE}$ is high less
energy is required to drive off the CE and thus it is easier to
eject the CE. This means that CE ejection occurs earlier in the
interaction with the result that most of the post-CE binaries will
systematically have relatively longer orbital periods. However, if
$\alpha_{\rm CE}$ is much lower than unity, then short-period CE
descendants along with mergers will be much more common. In Model
D with $\alpha_{\rm CE}=3$ the orbital periods of post-CE binaries
systematically increases so that close pre-CE binary systems that
would have coalesced during CE in Model A now survive and go on to
interact. On the other hand, post-CE binary systems that would
lead to mass-transfer in Model A now go through their entire
evolution independently. This variation in the evolutionary path
of post-CE binaries causes variations in the number of stars and
their distribution in the CMD for different evolutionary phases
and this leads to the observed discrepancies in the integrated
colours and ISEDs. We find that overall the number of binaries
experiencing RLOF does not change when $\alpha_{\rm CE}$ increases
but that the number experiencing more than one phase of RLOF
increases significantly.

In Figs \ref{ised-s-ace01} and \ref{ised-s-ace13} we give the
total flux, and the fluxes of all MS stars with mass $M \ga 0.7
{\rm M_\odot}$, CHeB stars and GB stars over the wavelength range
$4000 \le \lambda \le 10000 {\rm \AA}$ for Models A and D at ages
of 1\,Gyr and 13\,Gyr, respectively. Once again each flux curve is
expressed in units of the total flux at 2.2 $\mu$m (i.e., $F_{{\rm
\lambda,2.2} \mu {\rm m}}$) for the corresponding model. Fig.
\ref{ised-s-ace01} shows that the flux of all MS stars with mass
$M \ga 0.7 {\rm M_\odot}$ for Model D is bluer than that for Model
A, and there are no significant deviations in the GB and CHeB
fluxes. So the bluer ISEDs and integrated $\rm (U-B)$ and $\rm
(B-V)$ colours (see Figs \ref{del-ub} and \ref{del-bv}) for Model
D at $1\,$Gyr are mainly the result of the differences in MS stars
with mass $M \ga 0.7 {\rm M_\odot}$. For the $\tau=13$ \,Gyr BSPs,
Fig. \ref{ised-s-ace13} shows that the flux of all MS stars with
mass $M \ga 0.7 {\rm M_\odot}$ for Model D is also bluer than that
of Model A and the CHeB flux is redder. The net effect is the
bluer ISEDs and integrated $\rm (U-B)$ and $\rm (B-V)$ colours.

\subsection{Reimers' wind mass-loss efficiency  (Model E1-2)}
Mass-loss is one of the most important processes in stellar
astrophysics. Its value often determines the evolutionary path of
a star and therefore the final stages of a binary system. The
BSE/SSE packages of \citet{hur2000, hur2002} included a
prescription for mass loss which is not included in the detailed
models of \citet{pol98}. This prescription is drawn from a range
of current mass-loss theories available in the literature and can
easily be altered or added to (for details refer to Hurley et al.
2000).

We only discuss the influence of Reimers' empirical wind mass-loss
\citep{rei75} on our results because it is the most relevant among
these mass-loss mechanisms for populations of age $\tau \ge
1.0$\,Gyr. Reimers' wind mass-loss has been widely used in many
EPS studies owing to its simple parameterized form, and in the
SSE/BSE packages it is applied to the stellar envelope for
intermediate- and low-mass stars on the GB and beyond. The value
of the Reimers' mass-loss efficiency factor $\eta$ has been quoted
previously as varying from 0.25 to 2-3 \citep{dup86,kud78,ren81}.
The possibility of a metallicity dependence it is quite
controversial: it has been argued that there is a positive
metallicity dependence supported by observations and
hydrodynamical models \citep*{yi97}, while others have stated that
a metallicity dependence should not be included as there is no
strong evidence for it \citep{ibe83,car96}. In this study we do
not include a metallicity dependence in the Reimers' mass-loss,
which is to say that we adopt a fixed $\eta$ across all
metallicities in any particular model. In order to investigate the
effects of wind-loss on the model colours, three reasonable
choices for the Reimers' coefficient are used in equation
(\ref{wrei}): $\eta$= 0.0 (Model E1), 0.3 (Model A), and 0.5
(Model E2).

In Figs \ref{del-ub} and \ref{del-bv} the differences in the
integrated $\rm U-B$ and $\rm B-V$ colours between Model E1 and
Model A (i.e., $\Delta (\rm U-B)_{(E1-A)}$ and $\Delta (\rm
B-V)_{(E1-A)}$) and those between Model E2 and Model A (i.e.,
$\Delta (\rm U-B)_{(E2-A)}$ and $\Delta (\rm B-V)_{(E2-A)}$) are
given. By comparing the integrated $\rm U-B$ and $\rm B-V$ colours
for Models E1 and A we see that these two colours for Model E1
fluctuate around those for Model A at early age ($\tau \leq
8$\,Gyr), at intermediate and late ages the values of these two
colours for Model E1 are larger/redder than those for Model A, and
the discrepancies at intermediate and late ages are greater than
those at early age. Comparison of the integrated $\rm U-B$ and
$\rm B-V$ colours between Models E2 and A shows that there is no
systematic discrepancy in the two colours.

In summary, if neglecting the Reimers' mass-loss altogether (Model
E1) the integrated $\rm U-B$ and $\rm B-V$ colours are redder and
the populations look older at intermediate and late ages, while at
early age there is no systematic discrepancy ($\tau \leq 8$\,Gyr).
The variation of $\eta$ from 0.3 to 0.5 (Model E2) does not give
rise to systematic differences in the integrated $\rm U-B$ and
$\rm B-V$ colours.

The reason that the variation of $\eta$ can influence the
integrated colours, is that the variation in mass-loss owing to a
stellar wind can influence the spin orbital angular momentum of a
star and also the total orbital angular momentum. This is in
addition to the obvious effect it has on the stellar mass as well
as possibly the companion mass via accretion. So variation in
$\eta$ can alter the appearance of a star and even the evolution
path of a binary. This in turn will lead to a variation in the
distribution of stars in the CMD and the integrated colours,
especially in the extreme case of $\eta = 0$ (Model E1).

\subsection{The initial mass distribution of the secondaries  (Model F1-2)}
\begin{figure}
\psfig{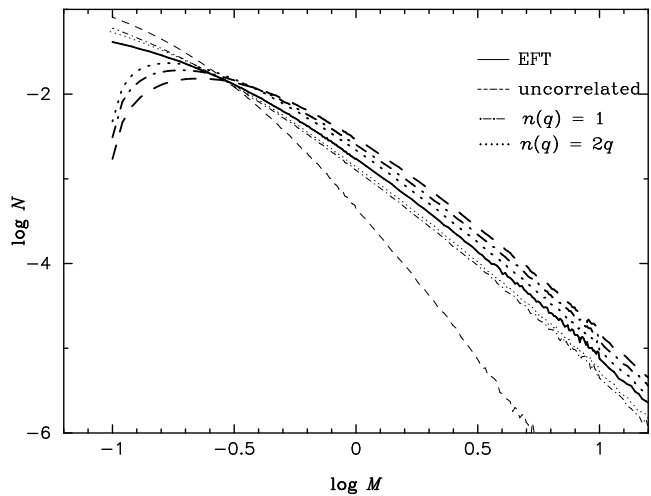}
\caption{The initial mass distributions of primaries (thick) and
secondaries (thin) for three assumptions: uncorrelated component
masses (Model F1), unique mass-ratio distribution $n(q)=1$ (Model
A) and thermal mass-ratio distribution $n(q)=2q$ (Model F2), and
the distribution of EFT (1989). Primary- and secondary-mass is in
the range $0.1 \le M \le 100 {\rm M_\odot}$.}
\label{figimf}
\end{figure}

The distribution of mass-ratio is less well-known than either of
the distributions over period or stellar mass. This is because
substantially more orbit data is required to determine a mass
ratio than for the total mass or period of a binary. Therefore, it
is quite controversial. In Model F1 we assume that the primary-
and secondary-masses are uncorrelated (both are chosen from
equation \ref{mdis}). \citet{duq91} found this to be an adequate
approximation for their sample of binaries whose primaries were
all F/G dwarfs like the Sun but it cannot be an adequate
approximation for massive stars in short-period ($P \la $25\,d,
\citealt{luc79}) and moderately-short-period ($P \le$ 3000\,d,
\citealt{maz92,tok92}) binaries. In Models F2 and A we assume that
the two component masses are correlated: Model A takes a constant
mass-ratio distribution (see equation \ref{qdis1}) while Model F2
takes a thermal form (see equation \ref{qdis2}).

From the discrepancy in the integrated $\rm (U-B)$ and $\rm (B-V)$
colours between Model F1 and Model A (i.e., $\Delta (\rm
U-B)_{(F1-A)}$ and $\Delta (\rm B-V)_{(F1-A)}$) and those between
Model F2 and Model A (i.e., $\Delta (\rm U-B)_{(F2-A)}$ and
$\Delta (\rm B-V)_{(F2-A)}$) in Figs \ref{del-ub} and
\ref{del-bv}, we see that the variation in the initial
secondary-mass distribution leads to fluctuations in the
integrated $\rm (U-B)$ and $\rm (B-V)$ colours and that the
fluctuation at late age is greater than that at early age.

In Fig. \ref{figimf} we present the initial mass distributions of
primaries and secondaries for Models A, F1 and F2. From it we see
that massive stars are more likely to have low-mass companions for
the distribution of uncorrelated component masses (Model F1) than
for those correlated cases (Models A and F2). Furthermore, the
possibility of massive stars paired with low-mass stars is greater
for a constant (Model A) than for a thermal initial mass-ratio
distribution (Model F2). In Fig. \ref{figimf} we also present the
distribution of EFT (1989, refer to equation \ref{mdis}). The
difference in the initial distribution leads to the discrepancy
observed in the integrated $\rm (U-B)$ and $\rm (B-V)$ colours.

\section{SUMMARY AND CONCLUSIONS}
We have simulated realistic stellar populations composed of 100\%
binaries by producing $1 \times 10^6$ binary systems using a Monte
Carlo technique. Using the EPS method we computed the integrated
colours, ISEDs and Lick/IDS absorption feature indices for an
extensive set of instantaneous burst BSPs with and without binary
interactions over a large range of age and metallicity: 1\,Gyr
$\le \tau \le$ 15\,Gyr and $-2.3 \le {\rm [Fe/H]} \le 0.2$. In our
EPS models we adopted the rapid SSE and BSE algorithms for the
single and binary evolution tracks, the empirical and
semi-empirical calibrated BaSeL-2.0 model for the library of
stellar spectra and empirical fitting functions for the Lick/IDS
spectral absorption feature indices. By comparing the results for
populations with and without binary interactions we show that the
inclusion of binary interactions makes each quantity we have
considered bluer and the appearance of the population younger or
the metallicity lower.

Also we have investigated the effect of input parameters (the CE
ejection efficiency $\alpha_{\rm CE}$ and the Reimers' wind
mass-loss coefficient $\eta$) and the input distributions
(eccentricity and the initial mass of the secondaries) on the
integrated $\rm U-B$ and $\rm B-V$ colours of a solar-metallicity
BSP. The results reveal that the variations in these
parameters/distributions can significantly affect the results.
Comparing the discrepancies in the integrated colours for all
models, we find that the differences between the models with and
without binary interactions are greater than those caused by the
variations in the choice of input parameters and distributions.
Based on the above results an important conclusion can be drawn
that it is very necessary to include binary interactions in EPS
studies.

The integrated colours and Lick/IDS absorption-line indices for
Models B-F, and the ISEDs for all models are not given for the
sake of the length of the paper but are available on request.

\section*{acknowledgements}
We acknowledge the generous support provided by the Chinese
Natural Science Foundation (Grant Nos 10303006, 19925312 \&
10273020), by the Chinese Academy of Sciences (KJCX2-SW-T06) and
by the 973 scheme (NKBRSF G1999075406). We are deeply indebted to
Dr. Lejeune for making his BaSeL-2.0 model available to us. We are
grateful to Dr L. Girardi, the referee, for his useful comments.

{}

\appendix
\onecolumn
\section[]{The absorption indices for Model A'}
\scriptsize
\begin{longtable}{rrrrrrrrrrrrrrrr}
 \hline
 \hline
 Age &
 $\scriptstyle 1 $&
 $\scriptstyle 2 $&
 $\scriptstyle 3 $&
 $\scriptstyle 4 $&
 $\scriptstyle 5 $&
 $\scriptstyle 6 $&
 $\scriptstyle 7 $&
 $\scriptstyle 8 $&
 $\scriptstyle 9 $&
 $\scriptstyle 10 $&
 $\scriptstyle 11 $ &
 $\scriptstyle 12 $&
 $\scriptstyle 13 $&
 $\scriptstyle 14 $&
 $\scriptstyle 15 $\\
 (Gyr)  \\
 \hline
 \endfirsthead
 \caption[]{continued}\\
 \hline
 \hline
 Age &
 $\scriptstyle 1 $&
 $\scriptstyle 2 $&
 $\scriptstyle 3 $&
 $\scriptstyle 4 $&
 $\scriptstyle 5 $&
 $\scriptstyle 6 $&
 $\scriptstyle 7 $&
 $\scriptstyle 8 $&
 $\scriptstyle 9 $&
 $\scriptstyle 10 $&
 $\scriptstyle 11 $&
 $\scriptstyle 12 $&
 $\scriptstyle 13 $&
 $\scriptstyle 14 $&
 $\scriptstyle 15 $\\
  (Gyr)  \\
\hline
 \endhead
 \multicolumn{16}{r}{continued on the next page}
 \endfoot
 \hline
 \endlastfoot
 \vspace*{1pt} \\
 \multicolumn{16}{c}{$z = 0.0001$} \\
${\rm CN_1}$        &  -0.228  &  -0.201  &  -0.178  &  -0.162  &  -0.163  &  -0.152  &  -0.152  &  -0.149  &  -0.146  &  -0.154  &  -0.147  &  -0.148  &  -0.149  &  -0.146  &  -0.147  \\
${\rm CN_2}$        &  -0.129  &  -0.116  &  -0.102  &  -0.095  &  -0.100  &  -0.093  &  -0.094  &  -0.091  &  -0.089  &  -0.096  &  -0.091  &  -0.092  &  -0.092  &  -0.089  &  -0.091  \\
Ca4227      &  -0.346  &  -0.215  &  -0.133  &  -0.052  &  -0.024  &   0.014  &   0.022  &   0.034  &   0.071  &   0.068  &   0.094  &   0.116  &   0.111  &   0.133  &   0.124  \\
G4300       &  -3.215  &  -2.326  &  -1.604  &  -1.097  &  -1.085  &  -0.758  &  -0.752  &  -0.648  &  -0.572  &  -0.756  &  -0.558  &  -0.545  &  -0.575  &  -0.468  &  -0.478  \\
Fe4383      &  -2.066  &  -1.555  &  -1.082  &  -0.821  &  -0.819  &  -0.611  &  -0.522  &  -0.421  &  -0.399  &  -0.506  &  -0.362  &  -0.353  &  -0.340  &  -0.280  &  -0.233  \\
Ca4455      &  -0.036  &   0.026  &   0.082  &   0.109  &   0.092  &   0.117  &   0.122  &   0.137  &   0.145  &   0.120  &   0.143  &   0.144  &   0.144  &   0.151  &   0.147  \\
Fe4531      &  -0.290  &  -0.014  &   0.195  &   0.304  &   0.241  &   0.334  &   0.345  &   0.386  &   0.407  &   0.308  &   0.391  &   0.383  &   0.388  &   0.395  &   0.366  \\
Fe4668      &   1.317  &   0.878  &   0.667  &   0.488  &   0.379  &   0.250  &   0.262  &   0.257  &   0.265  &   0.245  &   0.222  &   0.193  &   0.230  &   0.226  &   0.217  \\
${\rm H_\beta}$   &   5.888  &   4.963  &   4.320  &   3.896  &   3.948  &   3.706  &   3.682  &   3.583  &   3.485  &   3.639  &   3.426  &   3.393  &   3.370  &   3.259  &   3.262  \\
Fe5015      &   0.606  &   0.812  &   0.980  &   1.028  &   0.905  &   0.975  &   0.980  &   1.015  &   1.026  &   0.904  &   0.984  &   0.971  &   0.977  &   0.971  &   0.930  \\
${\rm Mg_1}$        &   0.011  &   0.005  &   0.003  &   0.002  &   0.001  &   0.000  &   0.001  &   0.001  &   0.003  &   0.004  &   0.004  &   0.005  &   0.007  &   0.008  &   0.009  \\
${\rm Mg_2}$        &   0.016  &   0.022  &   0.027  &   0.028  &   0.029  &   0.031  &   0.033  &   0.034  &   0.036  &   0.036  &   0.037  &   0.039  &   0.039  &   0.040  &   0.041  \\
${\rm Mg_b}$        &   0.373  &   0.495  &   0.525  &   0.559  &   0.731  &   0.769  &   0.790  &   0.856  &   0.820  &   0.871  &   0.851  &   0.871  &   0.875  &   0.887  &   0.907  \\
Fe5270      &  -0.807  &  -0.494  &  -0.270  &  -0.158  &  -0.253  &  -0.159  &  -0.149  &  -0.098  &  -0.075  &  -0.171  &  -0.073  &  -0.064  &  -0.055  &  -0.035  &  -0.059  \\
Fe5335      &   0.366  &   0.428  &   0.492  &   0.513  &   0.461  &   0.487  &   0.505  &   0.530  &   0.545  &   0.507  &   0.544  &   0.553  &   0.559  &   0.570  &   0.561  \\
Fe5406      &   0.107  &   0.148  &   0.195  &   0.205  &   0.153  &   0.173  &   0.184  &   0.204  &   0.215  &   0.176  &   0.209  &   0.212  &   0.219  &   0.224  &   0.214  \\
Fe5709      &   0.094  &   0.104  &   0.125  &   0.132  &   0.105  &   0.111  &   0.110  &   0.123  &   0.122  &   0.102  &   0.119  &   0.114  &   0.120  &   0.122  &   0.115  \\
Fe5782      &  -0.217  &  -0.187  &  -0.158  &  -0.148  &  -0.180  &  -0.169  &  -0.165  &  -0.153  &  -0.153  &  -0.180  &  -0.159  &  -0.163  &  -0.160  &  -0.160  &  -0.169  \\
Na D        &   1.149  &   1.050  &   1.095  &   1.002  &   0.853  &   0.886  &   0.911  &   0.934  &   0.951  &   0.902  &   0.932  &   0.999  &   0.997  &   1.001  &   0.997  \\
${\rm TiO_1}$       &   0.014  &   0.013  &   0.013  &   0.012  &   0.012  &   0.011  &   0.012  &   0.011  &   0.011  &   0.011  &   0.010  &   0.011  &   0.010  &   0.010  &   0.010  \\
${\rm TiO_2}$       &   0.000  &   0.000  &   0.000  &  -0.001  &  -0.003  &  -0.003  &  -0.002  &  -0.004  &  -0.003  &  -0.004  &  -0.004  &  -0.004  &  -0.004  &  -0.005  &  -0.005  \\

 \vspace*{1pt} \\
\multicolumn{16}{c}{$Z$ = 0.0003} \\

${\rm CN_1}$        &  -0.205  &  -0.181  &  -0.156  &  -0.137  &  -0.125  &  -0.111  &  -0.108  &  -0.105  &  -0.097  &  -0.099  &  -0.096  &  -0.101  &  -0.098  &  -0.106  &  -0.107  \\
${\rm CN_2}$        &  -0.117  &  -0.103  &  -0.087  &  -0.075  &  -0.070  &  -0.062  &  -0.060  &  -0.060  &  -0.053  &  -0.052  &  -0.052  &  -0.055  &  -0.051  &  -0.058  &  -0.058  \\
Ca4227      &  -0.191  &  -0.062  &   0.033  &   0.094  &   0.142  &   0.194  &   0.199  &   0.203  &   0.265  &   0.266  &   0.295  &   0.300  &   0.315  &   0.293  &   0.304  \\
G4300       &  -2.248  &  -1.395  &  -0.616  &  -0.011  &   0.318  &   0.775  &   0.902  &   0.998  &   1.235  &   1.139  &   1.236  &   1.157  &   1.205  &   0.972  &   0.975  \\
Fe4383      &  -1.431  &  -1.081  &  -0.672  &  -0.287  &  -0.152  &   0.150  &   0.306  &   0.319  &   0.528  &   0.618  &   0.640  &   0.632  &   0.689  &   0.572  &   0.571  \\
Ca4455      &  -0.058  &   0.065  &   0.148  &   0.217  &   0.225  &   0.269  &   0.280  &   0.272  &   0.310  &   0.317  &   0.323  &   0.320  &   0.345  &   0.309  &   0.311  \\
Fe4531      &   0.102  &   0.468  &   0.692  &   0.861  &   0.873  &   0.981  &   1.011  &   0.972  &   1.075  &   1.099  &   1.106  &   1.088  &   1.161  &   1.056  &   1.049  \\
Fe4668      &   0.092  &  -0.041  &  -0.121  &  -0.097  &  -0.221  &  -0.240  &  -0.260  &  -0.326  &  -0.301  &  -0.259  &  -0.278  &  -0.286  &  -0.224  &  -0.270  &  -0.280  \\
${\rm H_\beta}$     &   5.495  &   4.776  &   4.152  &   3.702  &   3.479  &   3.140  &   3.013  &   2.945  &   2.710  &   2.740  &   2.664  &   2.727  &   2.657  &   2.786  &   2.770  \\
Fe5015      &   0.735  &   1.146  &   1.383  &   1.542  &   1.476  &   1.571  &   1.599  &   1.499  &   1.626  &   1.676  &   1.651  &   1.636  &   1.757  &   1.609  &   1.585  \\
${\rm Mg_1}$        &   0.008  &   0.005  &   0.005  &   0.005  &   0.003  &   0.003  &   0.003  &   0.003  &   0.004  &   0.005  &   0.006  &   0.007  &   0.009  &   0.009  &   0.010  \\
${\rm Mg_2}$        &   0.029  &   0.037  &   0.042  &   0.047  &   0.041  &   0.046  &   0.048  &   0.049  &   0.050  &   0.050  &   0.051  &   0.054  &   0.056  &   0.053  &   0.053  \\
${\rm Mg_b}$       &   0.573  &   0.635  &   0.725  &   0.904  &   0.886  &   0.950  &   0.958  &   1.055  &   1.022  &   1.019  &   1.069  &   1.086  &   1.058  &   1.045  &   1.058  \\
Fe5270      &  -0.330  &   0.045  &   0.265  &   0.420  &   0.411  &   0.499  &   0.528  &   0.477  &   0.576  &   0.612  &   0.608  &   0.606  &   0.693  &   0.606  &   0.605  \\
Fe5335      &   0.273  &   0.422  &   0.524  &   0.597  &   0.554  &   0.603  &   0.620  &   0.585  &   0.640  &   0.661  &   0.667  &   0.678  &   0.734  &   0.685  &   0.683  \\
Fe5406      &   0.087  &   0.173  &   0.240  &   0.281  &   0.234  &   0.263  &   0.273  &   0.236  &   0.280  &   0.303  &   0.301  &   0.308  &   0.356  &   0.314  &   0.309  \\
Fe5709      &   0.127  &   0.173  &   0.216  &   0.243  &   0.243  &   0.252  &   0.253  &   0.238  &   0.260  &   0.277  &   0.270  &   0.261  &   0.288  &   0.271  &   0.269  \\
Fe5782      &  -0.075  &  -0.017  &   0.017  &   0.046  &   0.026  &   0.040  &   0.043  &   0.021  &   0.044  &   0.058  &   0.052  &   0.048  &   0.075  &   0.051  &   0.045  \\
Na D        &   1.052  &   1.067  &   1.086  &   1.095  &   0.874  &   0.911  &   0.949  &   0.831  &   0.896  &   0.962  &   0.917  &   1.018  &   1.118  &   1.007  &   0.989  \\
${\rm TiO_1}$       &   0.018  &   0.020  &   0.019  &   0.018  &   0.014  &   0.015  &   0.015  &   0.013  &   0.013  &   0.013  &   0.013  &   0.014  &   0.014  &   0.013  &   0.012  \\
${\rm TiO_2}$       &   0.009  &   0.014  &   0.013  &   0.013  &   0.004  &   0.005  &   0.006  &   0.001  &   0.003  &   0.002  &   0.002  &   0.004  &   0.006  &   0.002  &   0.001  \\

 \vspace*{1pt} \\
\multicolumn{16}{c}{$Z$ = 0.001} \\

${\rm CN_1}$        &  -0.192  &  -0.172  &  -0.136  &  -0.118  &  -0.101  &  -0.088  &  -0.081  &  -0.075  &  -0.076  &  -0.076  &  -0.071  &  -0.082  &  -0.086  &  -0.086  &  -0.098  \\
${\rm CN_2}$        &  -0.110  &  -0.097  &  -0.075  &  -0.062  &  -0.051  &  -0.043  &  -0.039  &  -0.035  &  -0.037  &  -0.038  &  -0.032  &  -0.041  &  -0.045  &  -0.045  &  -0.051  \\
Ca4227      &  -0.039  &   0.123  &   0.251  &   0.388  &   0.403  &   0.395  &   0.446  &   0.491  &   0.443  &   0.442  &   0.475  &   0.445  &   0.377  &   0.457  &   0.392  \\
G4300       &  -1.488  &  -0.791  &   0.417  &   1.122  &   1.654  &   2.053  &   2.294  &   2.488  &   2.468  &   2.454  &   2.599  &   2.283  &   2.161  &   2.208  &   1.759  \\
Fe4383      &  -0.955  &  -0.660  &  -0.003  &   0.494  &   0.828  &   1.136  &   1.209  &   1.365  &   1.341  &   1.332  &   1.422  &   1.416  &   1.341  &   1.380  &   1.341  \\
Ca4455      &   0.017  &   0.169  &   0.315  &   0.422  &   0.491  &   0.511  &   0.533  &   0.569  &   0.545  &   0.540  &   0.581  &   0.526  &   0.510  &   0.529  &   0.503  \\
Fe4531      &   0.522  &   0.880  &   1.234  &   1.448  &   1.563  &   1.624  &   1.657  &   1.731  &   1.673  &   1.666  &   1.762  &   1.630  &   1.605  &   1.643  &   1.606  \\
Fe4668      &  -0.356  &  -0.148  &  -0.171  &   0.084  &   0.143  &  -0.015  &   0.017  &   0.093  &  -0.019  &  -0.058  &   0.005  &  -0.082  &  -0.156  &  -0.158  &  -0.081  \\
${\rm H_\beta}$     &   5.269  &   4.684  &   3.893  &   3.415  &   3.080  &   2.797  &   2.629  &   2.488  &   2.496  &   2.474  &   2.369  &   2.519  &   2.639  &   2.585  &   2.834  \\
Fe5015      &   1.135  &   1.810  &   2.055  &   2.300  &   2.493  &   2.406  &   2.408  &   2.506  &   2.395  &   2.365  &   2.518  &   2.315  &   2.261  &   2.312  &   2.329  \\
${\rm Mg_1}$        &   0.015  &   0.014  &   0.013  &   0.016  &   0.015  &   0.015  &   0.016  &   0.018  &   0.017  &   0.017  &   0.019  &   0.018  &   0.017  &   0.020  &   0.021  \\
${\rm Mg_2}$        &   0.046  &   0.056  &   0.062  &   0.075  &   0.076  &   0.074  &   0.076  &   0.083  &   0.079  &   0.078  &   0.084  &   0.078  &   0.082  &   0.080  &   0.083  \\
${\rm Mg_b}$        &   0.798  &   1.106  &   1.176  &   1.388  &   1.574  &   1.391  &   1.442  &   1.497  &   1.447  &   1.417  &   1.403  &   1.447  &   1.616  &   1.448  &   1.464  \\
Fe5270      &   0.266  &   0.578  &   0.821  &   0.991  &   1.110  &   1.142  &   1.167  &   1.232  &   1.182  &   1.178  &   1.268  &   1.147  &   1.125  &   1.174  &   1.153  \\
Fe5335      &   0.410  &   0.611  &   0.726  &   0.917  &   0.934  &   0.935  &   0.943  &   1.024  &   0.937  &   0.935  &   1.003  &   0.918  &   0.891  &   0.926  &   0.946  \\
Fe5406      &   0.199  &   0.318  &   0.382  &   0.497  &   0.502  &   0.511  &   0.510  &   0.580  &   0.508  &   0.504  &   0.563  &   0.485  &   0.466  &   0.492  &   0.516  \\
Fe5709      &   0.230  &   0.274  &   0.338  &   0.337  &   0.381  &   0.402  &   0.411  &   0.423  &   0.415  &   0.420  &   0.447  &   0.408  &   0.406  &   0.416  &   0.412  \\
Fe5782      &   0.099  &   0.133  &   0.178  &   0.212  &   0.230  &   0.238  &   0.241  &   0.261  &   0.239  &   0.238  &   0.269  &   0.229  &   0.224  &   0.234  &   0.245  \\
Na D        &   1.169  &   1.147  &   1.086  &   1.283  &   1.224  &   1.159  &   1.156  &   1.268  &   1.190  &   1.142  &   1.266  &   1.167  &   1.156  &   1.255  &   1.286  \\
${\rm TiO_1}$       &   0.026  &   0.035  &   0.029  &   0.037  &   0.031  &   0.025  &   0.023  &   0.025  &   0.021  &   0.019  &   0.020  &   0.019  &   0.017  &   0.017  &   0.021  \\
${\rm TiO_2}$       &   0.025  &   0.042  &   0.031  &   0.048  &   0.037  &   0.025  &   0.021  &   0.026  &   0.017  &   0.013  &   0.018  &   0.014  &   0.010  &   0.011  &   0.019  \\

 \vspace*{1pt} \\
\multicolumn{16}{c}{$Z$ = 0.004} \\

${\rm CN_1}$        &  -0.185  &  -0.136  &  -0.105  &  -0.079  &  -0.064  &  -0.050  &  -0.047  &  -0.051  &  -0.038  &  -0.033  &  -0.031  &  -0.044  &  -0.037  &  -0.035  &  -0.035  \\
${\rm CN_2}$        &  -0.111  &  -0.077  &  -0.055  &  -0.036  &  -0.025  &  -0.014  &  -0.011  &  -0.014  &  -0.005  &  -0.001  &  -0.001  &  -0.010  &  -0.005  &  -0.004  &  -0.005  \\
Ca4227      &   0.159  &   0.339  &   0.427  &   0.518  &   0.604  &   0.652  &   0.696  &   0.690  &   0.710  &   0.792  &   0.789  &   0.775  &   0.778  &   0.856  &   0.803  \\
G4300       &  -0.955  &   0.661  &   1.690  &   2.564  &   3.081  &   3.502  &   3.658  &   3.581  &   4.012  &   4.258  &   4.316  &   3.978  &   4.207  &   4.366  &   4.359  \\
Fe4383      &  -0.406  &   0.375  &   1.101  &   1.677  &   2.017  &   2.322  &   2.429  &   2.482  &   2.684  &   2.873  &   2.878  &   2.820  &   3.020  &   3.098  &   3.156  \\
Ca4455      &   0.212  &   0.478  &   0.630  &   0.754  &   0.852  &   0.919  &   0.961  &   0.958  &   0.983  &   1.037  &   1.058  &   1.048  &   1.059  &   1.103  &   1.073  \\
Fe4531      &   1.040  &   1.553  &   1.825  &   2.055  &   2.202  &   2.327  &   2.379  &   2.352  &   2.429  &   2.524  &   2.536  &   2.501  &   2.558  &   2.597  &   2.570  \\
Fe4668      &  -0.096  &   0.409  &   0.800  &   1.030  &   1.355  &   1.460  &   1.639  &   1.703  &   1.506  &   1.592  &   1.716  &   1.822  &   1.701  &   1.891  &   1.678  \\
${\rm H_\beta}$     &   5.502  &   4.338  &   3.563  &   2.963  &   2.670  &   2.398  &   2.339  &   2.384  &   2.147  &   2.037  &   2.014  &   2.174  &   2.018  &   1.965  &   1.945  \\
Fe5015      &   2.102  &   2.967  &   3.313  &   3.473  &   3.815  &   3.878  &   4.098  &   4.168  &   3.874  &   3.977  &   4.146  &   4.300  &   3.996  &   4.243  &   3.952  \\
${\rm Mg_1}$        &   0.012  &   0.014  &   0.019  &   0.025  &   0.027  &   0.032  &   0.033  &   0.033  &   0.038  &   0.042  &   0.042  &   0.042  &   0.047  &   0.048  &   0.048  \\
${\rm Mg_2}$        &   0.064  &   0.080  &   0.092  &   0.101  &   0.112  &   0.117  &   0.123  &   0.124  &   0.124  &   0.133  &   0.136  &   0.140  &   0.146  &   0.147  &   0.144  \\
${\rm Mg_b}$        &   1.211  &   1.552  &   1.754  &   1.844  &   2.134  &   2.084  &   2.224  &   2.306  &   2.127  &   2.232  &   2.358  &   2.522  &   2.530  &   2.628  &   2.471  \\
Fe5270      &   0.829  &   1.246  &   1.480  &   1.652  &   1.789  &   1.886  &   1.953  &   1.940  &   1.963  &   2.035  &   2.060  &   2.076  &   2.088  &   2.143  &   2.110  \\
Fe5335      &   0.641  &   0.922  &   1.105  &   1.243  &   1.364  &   1.441  &   1.501  &   1.507  &   1.507  &   1.582  &   1.604  &   1.634  &   1.662  &   1.702  &   1.666  \\
Fe5406      &   0.296  &   0.492  &   0.617  &   0.723  &   0.793  &   0.868  &   0.888  &   0.883  &   0.918  &   0.969  &   0.971  &   0.974  &   1.020  &   1.027  &   1.013  \\
Fe5709      &   0.321  &   0.450  &   0.516  &   0.582  &   0.591  &   0.643  &   0.630  &   0.608  &   0.668  &   0.675  &   0.668  &   0.645  &   0.683  &   0.671  &   0.685  \\
Fe5782      &   0.214  &   0.299  &   0.353  &   0.406  &   0.422  &   0.462  &   0.461  &   0.447  &   0.475  &   0.495  &   0.486  &   0.476  &   0.509  &   0.499  &   0.502  \\
Na D        &   1.235  &   1.233  &   1.304  &   1.374  &   1.501  &   1.510  &   1.615  &   1.651  &   1.598  &   1.769  &   1.726  &   1.802  &   1.843  &   1.893  &   1.893  \\
${\rm TiO_1}$       &   0.035  &   0.035  &   0.035  &   0.031  &   0.037  &   0.033  &   0.039  &   0.043  &   0.031  &   0.035  &   0.038  &   0.043  &   0.035  &   0.039  &   0.033  \\
${\rm TiO_2}$       &   0.041  &   0.042  &   0.041  &   0.036  &   0.047  &   0.041  &   0.053  &   0.059  &   0.038  &   0.046  &   0.051  &   0.061  &   0.047  &   0.054  &   0.042  \\

 \vspace*{1pt} \\
\multicolumn{16}{c}{$Z$ = 0.01} \\

${\rm CN_1}$        &  -0.155  &  -0.087  &  -0.056  &  -0.045  &  -0.036  &  -0.021  &  -0.022  &  -0.012  &  -0.005  &  -0.006  &  -0.003  &  -0.002  &   0.001  &   0.006  &   0.004  \\
${\rm CN_2}$        &  -0.092  &  -0.041  &  -0.017  &  -0.007  &   0.000  &   0.013  &   0.014  &   0.022  &   0.028  &   0.027  &   0.028  &   0.030  &   0.032  &   0.035  &   0.034  \\
Ca4227      &   0.276  &   0.518  &   0.644  &   0.695  &   0.720  &   0.820  &   0.826  &   0.875  &   0.919  &   0.942  &   0.948  &   0.955  &   0.968  &   1.007  &   0.992  \\
G4300       &   0.051  &   2.221  &   3.186  &   3.524  &   3.800  &   4.221  &   4.179  &   4.450  &   4.672  &   4.707  &   4.808  &   4.820  &   4.913  &   5.121  &   5.076  \\
Fe4383      &   0.379  &   1.805  &   2.575  &   2.912  &   3.302  &   3.579  &   3.647  &   3.839  &   4.027  &   4.082  &   4.149  &   4.228  &   4.313  &   4.400  &   4.422  \\
Ca4455      &   0.463  &   0.828  &   0.994  &   1.067  &   1.127  &   1.213  &   1.218  &   1.278  &   1.325  &   1.340  &   1.352  &   1.364  &   1.389  &   1.421  &   1.425  \\
Fe4531      &   1.496  &   2.148  &   2.430  &   2.545  &   2.646  &   2.784  &   2.783  &   2.885  &   2.959  &   2.977  &   2.986  &   3.010  &   3.048  &   3.076  &   3.091  \\
Fe4668      &   0.613  &   1.795  &   2.304  &   2.544  &   2.711  &   2.971  &   2.985  &   3.137  &   3.252  &   3.279  &   3.290  &   3.291  &   3.339  &   3.415  &   3.382  \\
${\rm H_\beta}$     &   5.133  &   3.487  &   2.836  &   2.621  &   2.475  &   2.254  &   2.270  &   2.109  &   1.995  &   1.976  &   1.930  &   1.897  &   1.854  &   1.774  &   1.767  \\
Fe5015      &   2.768  &   3.745  &   4.070  &   4.245  &   4.342  &   4.494  &   4.508  &   4.613  &   4.686  &   4.733  &   4.750  &   4.730  &   4.784  &   4.874  &   4.863  \\
${\rm Mg_1}$        &   0.014  &   0.026  &   0.037  &   0.041  &   0.047  &   0.053  &   0.054  &   0.060  &   0.064  &   0.065  &   0.065  &   0.069  &   0.070  &   0.071  &   0.074  \\
${\rm Mg_2}$        &   0.076  &   0.111  &   0.129  &   0.138  &   0.147  &   0.159  &   0.162  &   0.169  &   0.177  &   0.181  &   0.182  &   0.185  &   0.189  &   0.191  &   0.194  \\
${\rm Mg_b}$        &   1.323  &   1.961  &   2.137  &   2.299  &   2.388  &   2.546  &   2.592  &   2.676  &   2.764  &   2.850  &   2.896  &   2.905  &   2.952  &   3.042  &   3.020  \\
Fe5270      &   1.262  &   1.815  &   2.053  &   2.160  &   2.247  &   2.356  &   2.372  &   2.449  &   2.507  &   2.534  &   2.547  &   2.565  &   2.595  &   2.622  &   2.639  \\
Fe5335      &   0.963  &   1.450  &   1.669  &   1.765  &   1.850  &   1.956  &   1.962  &   2.041  &   2.099  &   2.122  &   2.127  &   2.150  &   2.178  &   2.193  &   2.213  \\
Fe5406      &   0.510  &   0.840  &   1.009  &   1.079  &   1.146  &   1.228  &   1.237  &   1.297  &   1.343  &   1.348  &   1.352  &   1.378  &   1.394  &   1.396  &   1.414  \\
Fe5709      &   0.513  &   0.655  &   0.737  &   0.769  &   0.794  &   0.825  &   0.828  &   0.853  &   0.869  &   0.862  &   0.865  &   0.881  &   0.884  &   0.877  &   0.889  \\
Fe5782      &   0.346  &   0.491  &   0.561  &   0.590  &   0.616  &   0.649  &   0.652  &   0.676  &   0.692  &   0.689  &   0.683  &   0.695  &   0.700  &   0.689  &   0.699  \\
Na D        &   1.256  &   1.606  &   1.789  &   1.879  &   1.976  &   2.086  &   2.153  &   2.186  &   2.257  &   2.338  &   2.347  &   2.355  &   2.390  &   2.431  &   2.457  \\
${\rm TiO_1}$       &   0.027  &   0.031  &   0.030  &   0.031  &   0.032  &   0.034  &   0.034  &   0.034  &   0.034  &   0.036  &   0.035  &   0.033  &   0.034  &   0.036  &   0.036  \\
${\rm TiO_2}$       &   0.024  &   0.035  &   0.036  &   0.038  &   0.040  &   0.044  &   0.044  &   0.045  &   0.047  &   0.050  &   0.047  &   0.044  &   0.047  &   0.049  &   0.049  \\

 \vspace*{1pt} \\
\multicolumn{16}{c}{$Z$ = 0.02} \\

${\rm CN_1}$        &  -0.121  &  -0.061  &  -0.030  &  -0.011  &   0.004  &   0.015  &   0.027  &   0.020  &   0.033  &   0.046  &   0.048  &   0.042  &   0.056  &   0.030  &   0.063  \\
${\rm CN_2}$        &  -0.065  &  -0.018  &   0.008  &   0.025  &   0.038  &   0.048  &   0.060  &   0.055  &   0.065  &   0.077  &   0.080  &   0.074  &   0.089  &   0.068  &   0.095  \\
Ca4227      &   0.411  &   0.690  &   0.834  &   0.917  &   0.994  &   1.030  &   1.126  &   1.096  &   1.136  &   1.238  &   1.256  &   1.186  &   1.286  &   1.177  &   1.363  \\
G4300       &   0.995  &   2.939  &   3.783  &   4.214  &   4.592  &   4.879  &   5.132  &   4.962  &   5.284  &   5.538  &   5.543  &   5.458  &   5.702  &   5.024  &   5.845  \\
Fe4383      &   1.294  &   2.778  &   3.674  &   4.146  &   4.497  &   4.790  &   5.077  &   5.091  &   5.313  &   5.555  &   5.635  &   5.668  &   5.833  &   5.569  &   6.059  \\
Ca4455      &   0.739  &   1.083  &   1.262  &   1.365  &   1.452  &   1.516  &   1.591  &   1.575  &   1.639  &   1.714  &   1.729  &   1.705  &   1.774  &   1.683  &   1.831  \\
Fe4531      &   1.934  &   2.531  &   2.821  &   2.990  &   3.117  &   3.217  &   3.329  &   3.311  &   3.400  &   3.509  &   3.529  &   3.502  &   3.598  &   3.463  &   3.687  \\
Fe4668      &   1.873  &   3.086  &   3.765  &   4.128  &   4.431  &   4.659  &   4.922  &   4.844  &   5.092  &   5.324  &   5.383  &   5.213  &   5.408  &   5.233  &   5.588  \\
${\rm H_\beta}$     &   4.655  &   3.137  &   2.623  &   2.371  &   2.183  &   2.040  &   1.899  &   1.962  &   1.795  &   1.687  &   1.645  &   1.676  &   1.558  &   1.795  &   1.465  \\
Fe5015      &   3.484  &   4.316  &   4.732  &   4.909  &   5.113  &   5.215  &   5.376  &   5.290  &   5.444  &   5.611  &   5.629  &   5.503  &   5.681  &   5.506  &   5.714  \\
${\rm Mg_1}$        &   0.021  &   0.043  &   0.057  &   0.067  &   0.072  &   0.078  &   0.085  &   0.086  &   0.091  &   0.095  &   0.098  &   0.097  &   0.101  &   0.100  &   0.109  \\
${\rm Mg_2}$        &   0.099  &   0.140  &   0.167  &   0.183  &   0.195  &   0.207  &   0.218  &   0.219  &   0.229  &   0.238  &   0.243  &   0.239  &   0.247  &   0.244  &   0.258  \\
${\rm Mg_b}$        &   1.635  &   2.202  &   2.592  &   2.803  &   3.004  &   3.224  &   3.321  &   3.305  &   3.590  &   3.593  &   3.641  &   3.572  &   3.629  &   3.616  &   3.795  \\
Fe5270      &   1.691  &   2.204  &   2.464  &   2.610  &   2.728  &   2.804  &   2.901  &   2.887  &   2.967  &   3.052  &   3.082  &   3.054  &   3.158  &   3.060  &   3.202  \\
Fe5335      &   1.420  &   1.914  &   2.163  &   2.305  &   2.407  &   2.485  &   2.580  &   2.572  &   2.650  &   2.725  &   2.752  &   2.725  &   2.800  &   2.736  &   2.868  \\
Fe5406      &   0.798  &   1.149  &   1.335  &   1.452  &   1.524  &   1.583  &   1.654  &   1.654  &   1.706  &   1.761  &   1.784  &   1.769  &   1.832  &   1.776  &   1.882  \\
Fe5709      &   0.674  &   0.828  &   0.905  &   0.956  &   0.982  &   1.001  &   1.025  &   1.022  &   1.029  &   1.048  &   1.052  &   1.053  &   1.076  &   1.041  &   1.083  \\
Fe5782      &   0.501  &   0.647  &   0.721  &   0.770  &   0.796  &   0.817  &   0.843  &   0.841  &   0.856  &   0.876  &   0.883  &   0.874  &   0.901  &   0.875  &   0.912  \\
Na D        &   1.633  &   2.146  &   2.400  &   2.569  &   2.705  &   2.791  &   2.913  &   2.941  &   3.040  &   3.133  &   3.196  &   3.191  &   3.329  &   3.275  &   3.415  \\
${\rm TiO_1}$       &   0.023  &   0.027  &   0.029  &   0.030  &   0.032  &   0.033  &   0.034  &   0.034  &   0.038  &   0.039  &   0.039  &   0.035  &   0.037  &   0.039  &   0.038  \\
${\rm TiO_2}$       &   0.018  &   0.029  &   0.037  &   0.039  &   0.042  &   0.045  &   0.049  &   0.048  &   0.056  &   0.058  &   0.059  &   0.051  &   0.056  &   0.058  &   0.058  \\

 \vspace*{1pt} \\
\multicolumn{16}{c}{$Z$ = 0.03} \\

${\rm CN_1}$&  -0.104  &  -0.042  &  -0.020  &   0.001  &   0.015  &   0.026  &   0.030  &   0.041  &   0.054  &   0.055  &   0.052  &   0.074  &   0.091  &   0.087  &   0.082  \\
${\rm CN_2}$&  -0.051  &  -0.003  &   0.017  &   0.035  &   0.048  &   0.060  &   0.064  &   0.075  &   0.087  &   0.088  &   0.089  &   0.108  &   0.126  &   0.122  &   0.117  \\
Ca4227      &   0.514  &   0.806  &   0.940  &   1.081  &   1.145  &   1.232  &   1.272  &   1.354  &   1.459  &   1.414  &   1.438  &   1.528  &   1.670  &   1.689  &   1.596  \\
G4300       &   1.524  &   3.520  &   4.050  &   4.534  &   4.870  &   5.021  &   5.115  &   5.276  &   5.584  &   5.560  &   5.373  &   5.816  &   5.986  &   6.004  &   5.868  \\
Fe4383      &   1.768  &   3.522  &   4.333  &   4.772  &   5.142  &   5.429  &   5.617  &   5.855  &   6.084  &   6.214  &   6.267  &   6.520  &   6.830  &   6.828  &   6.729  \\
Ca4455      &   0.910  &   1.265  &   1.416  &   1.538  &   1.617  &   1.681  &   1.720  &   1.787  &   1.873  &   1.871  &   1.884  &   1.967  &   2.042  &   2.049  &   2.031  \\
Fe4531      &   2.180  &   2.785  &   3.035  &   3.227  &   3.356  &   3.453  &   3.514  &   3.618  &   3.742  &   3.746  &   3.780  &   3.894  &   4.011  &   4.008  &   3.989  \\
Fe4668      &   2.735  &   4.025  &   4.650  &   5.102  &   5.347  &   5.634  &   5.739  &   6.045  &   6.320  &   6.255  &   6.316  &   6.582  &   6.826  &   6.840  &   6.778  \\
${\rm H_\beta}$   &   4.351  &   2.879  &   2.553  &   2.268  &   2.078  &   1.941  &   1.882  &   1.759  &   1.646  &   1.614  &   1.645  &   1.465  &   1.345  &   1.333  &   1.381  \\
Fe5015      &   3.958  &   4.731  &   5.076  &   5.334  &   5.428  &   5.567  &   5.600  &   5.769  &   5.973  &   5.901  &   5.912  &   6.047  &   6.157  &   6.190  &   6.151  \\
${\rm Mg_1}$&   0.029  &   0.056  &   0.072  &   0.083  &   0.092  &   0.101  &   0.105  &   0.113  &   0.117  &   0.120  &   0.125  &   0.130  &   0.139  &   0.137  &   0.137  \\
${\rm Mg_2}$&   0.114  &   0.162  &   0.192  &   0.212  &   0.227  &   0.240  &   0.247  &   0.261  &   0.272  &   0.272  &   0.280  &   0.293  &   0.301  &   0.303  &   0.308  \\
${\rm Mg_b}$&   1.783  &   2.440  &   2.881  &   3.183  &   3.411  &   3.525  &   3.615  &   3.791  &   3.968  &   3.942  &   3.991  &   4.253  &   4.221  &   4.351  &   4.488  \\
Fe5270      &   1.943  &   2.451  &   2.683  &   2.845  &   2.946  &   3.044  &   3.091  &   3.185  &   3.274  &   3.286  &   3.329  &   3.405  &   3.505  &   3.504  &   3.492  \\
Fe5335      &   1.732  &   2.238  &   2.472  &   2.636  &   2.741  &   2.840  &   2.888  &   2.983  &   3.066  &   3.077  &   3.121  &   3.188  &   3.278  &   3.278  &   3.264  \\
Fe5406      &   0.981  &   1.353  &   1.529  &   1.650  &   1.736  &   1.812  &   1.852  &   1.922  &   1.975  &   1.991  &   2.031  &   2.083  &   2.162  &   2.143  &   2.129  \\
Fe5709      &   0.772  &   0.943  &   1.011  &   1.048  &   1.086  &   1.108  &   1.121  &   1.131  &   1.139  &   1.162  &   1.172  &   1.187  &   1.219  &   1.194  &   1.184  \\
Fe5782      &   0.588  &   0.734  &   0.809  &   0.854  &   0.887  &   0.917  &   0.927  &   0.956  &   0.968  &   0.978  &   0.999  &   1.013  &   1.051  &   1.025  &   1.024  \\
Na D        &   1.937  &   2.470  &   2.783  &   3.002  &   3.138  &   3.291  &   3.370  &   3.524  &   3.626  &   3.650  &   3.752  &   3.833  &   4.003  &   4.017  &   4.027  \\
${\rm TiO_1}$     &   0.024  &   0.024  &   0.029  &   0.033  &   0.032  &   0.035  &   0.034  &   0.040  &   0.043  &   0.037  &   0.039  &   0.039  &   0.040  &   0.041  &   0.042  \\
${\rm TiO_2}$     &   0.021  &   0.027  &   0.038  &   0.048  &   0.047  &   0.053  &   0.053  &   0.065  &   0.070  &   0.060  &   0.064  &   0.065  &   0.067  &   0.068  &   0.070  \\
\label{licktab}
\end{longtable}

\bsp
\label{lastpage}
\end{document}